\documentclass[manuscript]{acmart}

\usepackage{acronym} 
\usepackage{algorithmic}
\usepackage{booktabs}
\usepackage{clrscode3e}
\usepackage{multirow}
\usepackage{graphicx}
\usepackage{subfigure} 
\usepackage{textcomp}
\usepackage[table]{xcolor}
\usepackage{tabularx}
\usepackage{xspace}
\usepackage[normalem]{ulem}
\usepackage{varwidth}
\usepackage{paralist}
\usepackage{wrapfig}
\usepackage{bm}
\usepackage{enumitem}
\usepackage{placeins}
\usepackage{tikz}
\useunder{\uline}{\ul}{}
\usepackage[ruled,vlined,linesnumbered]{algorithm2e}
\usepackage{threeparttable}
\usepackage{url}
\usepackage{bibentry}

\acrodef{MoE}{Mixture-of-Experts}
\acrodef{LLM}{Large Language Model}
\acrodef{LLMs}{Large Language Models}
\acrodef{CDR}{Cross-Domain Recommendation}
\acrodef{MHA}{multi-head self-attention layer}

\AtBeginDocument{%
  \providecommand\BibTeX{{%
    \normalfont B\kern-0.5em{\scshape i\kern-0.25em b}\kern-0.8em\TeX}}}

\setcopyright{acmcopyright}
\copyrightyear{2027}
\acmYear{2027}
\acmDOI{XXXXXXX.XXXXXXX}
\acmJournal{TOIS}
\acmVolume{0}
\acmNumber{0}
\acmArticle{0}
\acmMonth{0}
\begin{document}

\title{B\'ezier-Guided Adaptive Flow for Cross-domain Sequential Recommendation}
\author{Ziang Lu}
\email{zianglu@outlook.com}
\affiliation{
  \institution{Anhui University}
  \city{Hefei}
  \country{China}
}

\author{Yu Zhang}
\email{zhangyu.ahu@gmail.com}
\affiliation{
  \institution{Anhui University}
  \city{Hefei}
  \country{China}
}

\author{Xinzhe Jiang}
\email{jiangxinzhe9@gmail.com}
\affiliation{
  \institution{Anhui University}
  \city{Hefei}
  \country{China}
}

\author{Lei Sang}
\email{sanglei@ahu.edu.cn}
\affiliation{
  \institution{Anhui University}
  \city{Hefei}
  \country{China}
}

\author{Yiwen Zhang}
\authornote{Corresponding author.}
\email{zhangyiwen@ahu.edu.cn}

\affiliation{
  \institution{Anhui University}
  \city{Hefei}
  \country{China}
}

\begin{abstract}
Cross-domain sequential recommendation aims to alleviate data sparsity and cold-start problems by transferring useful preference knowledge across domains. However, users exhibit varying degrees of preference consistency across domains, while existing methods often rely on global alignment objectives or coarse-grained transfer strategies. Such approaches may introduce irrelevant information and provide limited flexibility in adapting semantic representations to individual user preferences. To address these challenges, we propose BGAFlow, a personalized framework that combines state-adaptive semantic evolution with user-specific transfer control. BGAFlow first establishes a shared semantic space through unified dual-domain pretraining with user profiles inferred by a large language model (LLM). Based on latent interest overlap, it then estimates a personalized horizon to determine the appropriate transfer extent for each user. To enable semantic adaptation, BGAFlow constructs a conditional Bézier curve as a geometric reference and learns a residual flow that progressively refines the trajectory based on intermediate preference states. The learned flow incorporates dynamic preference feedback to adjust transfer directions, while recommendation supervision and trajectory regularization encourage effective and smooth adaptation. Extensive experiments on real-world datasets demonstrate that BGAFlow consistently outperforms state-of-the-art baselines, validating the benefits of personalized transfer control and state-adaptive semantic evolution.
\end{abstract}


\maketitle

\section{Introduction}

Cross-domain recommendation (CDR) aims to improve recommendation quality by leveraging user behaviors from multiple related domains~\cite{zang2022survey, chen2024survey}. In many real-world platforms, users interact with heterogeneous services, such as movies, books, music, news, and e-commerce items, and their behaviors in different domains may reveal complementary aspects of their preferences. By transferring useful knowledge across domains, CDR has been widely studied as an effective solution to data sparsity and cold-start problems~\cite{li2024cdrnp, li2025disco, zhao2024domain}. This problem is particularly important in sequential recommendation, where models need to capture users' evolving interests from historical interaction sequences~\cite{kang2018self, boka2024survey, liu2026bridging} while exploiting transferable preference signals from other domains.

\begin{figure}
    \centering
    \includegraphics[width=8.5cm]{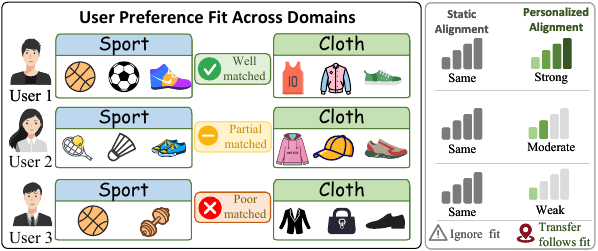}
    \caption{Motivation for personalized cross-domain transfer. Users exhibit different degrees of cross-domain preference consistency, requiring user-specific control over the extent of knowledge transfer.}
    \Description{Three example users have well-matched, partially matched, or poorly matched interests across Sport and Cloth. Personalized alignment assigns strong, moderate, or weak transfer to these cases.}
\label{fig:moti}
\end{figure}

Existing CDR methods commonly transfer knowledge through shared or parallel architectures~\cite{hu2018conet, guo2021gcn, ma2019pi}, global alignment objectives such as contrastive or adversarial learning~\cite{cao2022contrastive, zhao2023cross, ma2024triple, lu2023contrastive, zhang2021deep, guo2023dan, su2024cross, li2022adversarial}, domain-general or invariant representation learning~\cite{li2022recguru, liu2022exploiting, bian2025abxi, huang2024diit}, and cooperative transfer-control strategies~\cite{park2024pacer, park2023cracking, song2024mitigating, ju2025revisiting}. Although these methods have achieved promising performance, many still rely on predefined representation mappings, global alignment objectives, or coarse-grained transfer strategies, offering limited flexibility in adapting the transfer process to individual users. In practice, users exhibit different degrees of preference consistency across domains and may require different amounts of transferred knowledge. Moreover, cross-domain preference adaptation may involve gradual adjustments among related semantic interests rather than simple representation fusion. As illustrated in Fig.~\ref{fig:moti}, some users have well-matched interests, some are only partially matched, and others show weak cross-domain correspondence. Applying similar transfer strengths or inflexible adaptation mechanisms to these users may introduce irrelevant information from the paired domain, resulting in negative transfer.

\begin{figure}
    \centering
    \subfigure[Interest similarity distribution]{
        \includegraphics[width=0.4\linewidth]{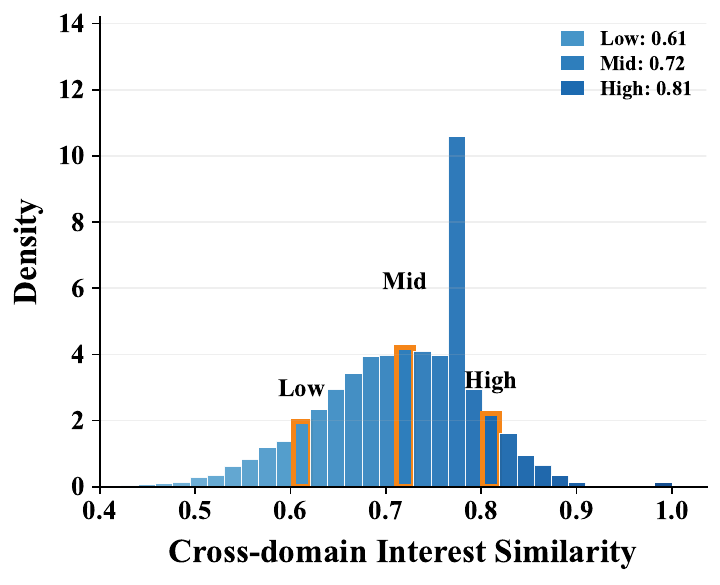}
        \label{fig:intro_interest}
    }
    \hfill
    \subfigure[Alignment strength comparison]{
        \includegraphics[width=0.4\linewidth]{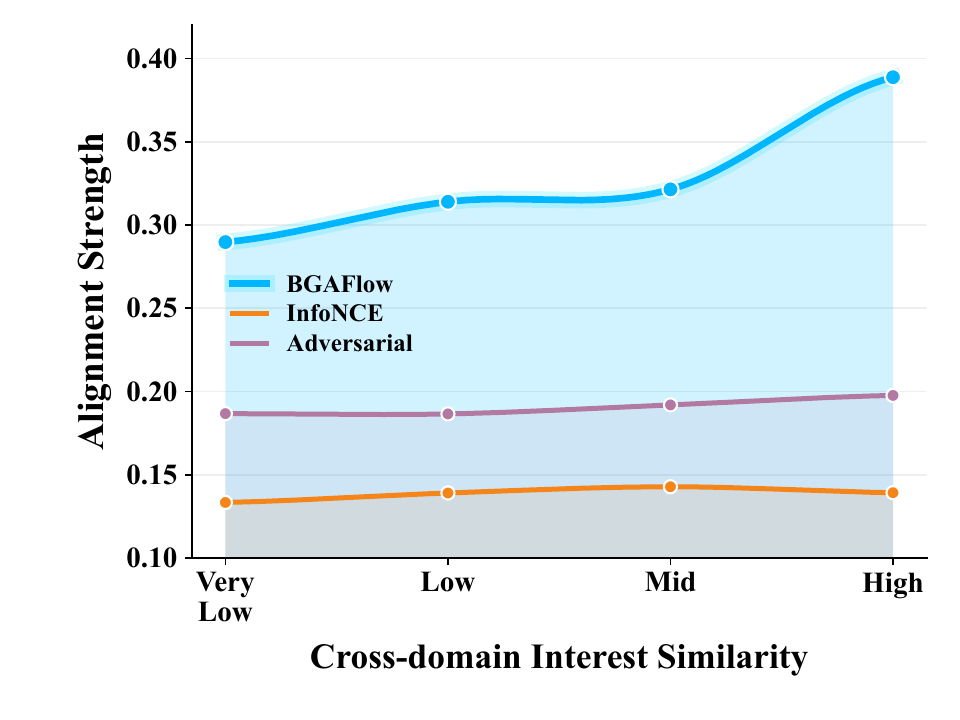}
        \label{fig:intro_align}
    }
    \caption{Cross-domain interest similarity and alignment strength on Sport-Cloth. \ac{LLM}-inferred user interests exhibit diverse cross-domain similarity, while BGAFlow exhibits greater variation in alignment strength across user groups than InfoNCE and adversarial alignment.}
    \Description{The left panel shows a distribution of cross-domain interest similarity on Sport-Cloth. The right panel compares BGAFlow, InfoNCE, and adversarial alignment across four similarity groups. BGAFlow increases its alignment strength more clearly as interest similarity increases.}
\label{fig:intro_motivation}
\end{figure}

To further examine these differences in real cross-domain data, we conduct an empirical analysis on the Sport-Cloth dataset. Specifically, we first use LLM-inferred user interests to measure cross-domain preference similarity. As shown in Fig.~\ref{fig:intro_interest}, users exhibit substantial variations in preference consistency, indicating different degrees of transferable interests. We then compare the alignment strengths of our proposed BGAFlow with those of InfoNCE and adversarial alignment. Fig.~\ref{fig:intro_align} shows that the two conventional objectives yield relatively stable alignment strengths across similarity groups, whereas BGAFlow produces stronger alignment for users with higher interest similarity. This observation highlights the need for personalized transfer control rather than uniform alignment. Beyond determining the transfer extent, effective cross-domain adaptation also requires a mechanism that can progressively refine semantic representations according to their intermediate states. These considerations raise two key challenges:

\begin{itemize}[leftmargin=*]
    \item \textbf{State-adaptive semantic evolution.} Cross-domain transfer requires a mechanism that progressively adapts semantic representations toward relevant preferences in the paired domain, rather than relying solely on one-step fusion or direct mapping. More importantly, the transfer process should respond to intermediate semantic states and adjust subsequent directions accordingly, while maintaining a coherent and smooth evolution trajectory.

    \item \textbf{Reliable user-specific alignment-strength estimation.} To control the transfer extent, the model needs a semantically meaningful signal that reflects how strongly a user's interests are related across two domains. Such a signal should capture user-level transferable preference overlap and determine the appropriate alignment strength, while avoiding unreliable direct comparisons between heterogeneous domain-specific representations.
\end{itemize}

To address these challenges, we propose B\'ezier-Guided Adaptive Flow (BGAFlow), which combines state-adaptive semantic refinement with personalized transfer control. BGAFlow first performs unified dual-domain pretraining to learn comparable semantic representations while preserving domain-specific collaborative information. It then estimates cross-domain preference overlap through latent interest prototypes to determine a personalized transfer horizon. For semantic adaptation, BGAFlow constructs a conditional B\'ezier curve as a geometric reference and learns a residual flow that progressively refines the trajectory based on intermediate preference states. Dynamic prototype feedback enables the flow to adjust its correction directions during transfer, while task-guided flow matching provides recommendation-aware supervision. Furthermore, smoothstep-based integration and trajectory regularization encourage gradual and stable representation evolution. Together, these designs enable BGAFlow to adapt both the extent and the process of cross-domain knowledge transfer to individual user preferences.

Our contributions are summarized as follows:
\begin{itemize}[leftmargin=*]
    \item We propose BGAFlow, a personalized framework for cross-domain sequential recommendation that combines user-specific transfer control with state-adaptive semantic evolution.

    \item We develop a B\'ezier-guided residual flow that progressively refines a structured geometric reference using dynamic preference feedback and task-guided velocity supervision. Smoothstep-based integration and trajectory regularization further promote stable semantic adaptation.

    \item We conduct extensive experiments on real-world cross-domain recommendation datasets, demonstrating the effectiveness of BGAFlow against state-of-the-art baselines and validating its key components through comprehensive analyses.
\end{itemize}

\section{Related Work}

\subsection{Cross-domain Recommendation}

Cross-domain recommendation (CDR) aims to alleviate data sparsity by transferring complementary preference knowledge across domains. Early studies mainly establish cross-domain connections through shared parameters or parallel network architectures. For example, CoNet transfers hidden representations between domain-specific networks, while $\pi$-Net exchanges sequential preference information through parallel information-sharing modules~\cite{hu2018conet,ma2019pi}. DA-GCN further employs domain-aware graph propagation to capture cross-domain relationships under shared accounts~\cite{guo2021gcn}. These approaches facilitate knowledge sharing between domains, but their transfer processes are largely determined by predefined architectural connections.

Another line of research focuses on learning transferable representations through cross-domain alignment. C$^2$DSR and Tri-CDR exploit sequential correlations through contrastive learning and multi-sequence modeling~\cite{cao2022contrastive,ma2024triple}. Other studies employ contrastive or adversarial objectives to encourage representation consistency across domains~\cite{zhao2023cross,guo2023dan,su2024cross,lu2023contrastive}. Meanwhile, RecGURU, DIIT, and ABXI seek to extract domain-general or invariant preference information to improve knowledge transfer~\cite{li2022recguru,huang2024diit,bian2025abxi}. Recognizing that cross-domain signals are not universally beneficial, recent studies have also explored cooperative learning and transferability estimation to mitigate negative transfer~\cite{park2023cracking,park2024pacer,song2024mitigating,ju2025revisiting}. Despite these advances, user-specific transfer strength and intermediate semantic adaptation are not always explicitly modeled together. In contrast, BGAFlow combines personalized transfer control with a learned semantic evolution process, allowing intermediate representations to be progressively refined according to their current preference states.

\subsection{LLM-based Recommendation}

Large language models (LLMs) have recently attracted considerable attention in recommendation due to their semantic understanding and generalization capabilities~\cite{wu2024survey}. Existing studies explore their potential through task adaptation and semantic enhancement. TALLRec introduces an efficient tuning framework to align language models with recommendation tasks~\cite{bao2023tallrec}. LLMRec leverages LLM-generated information to augment user--item interaction graphs, while LLM-ESR integrates LLM-derived semantic embeddings with collaborative signals to improve long-tailed sequential recommendation~\cite{wei2024llmrec,liu2024llm}. These methods demonstrate that language-based knowledge can complement behavioral signals and enrich user preference representations.

More recently, LLMs have been incorporated into cross-domain sequential recommendation to bridge semantic differences between domains~\cite{qin2026bridging}. LeCDSR utilizes LLM-generated semantic embeddings and cross-domain user profiles to facilitate preference transfer~\cite{wang2025lecdsr}. LLM4CDSR combines unified item representations with hierarchical user profiling to model cross-domain preferences, while LLMCDSR investigates LLM-enhanced knowledge transfer involving overlapping and non-overlapping users~\cite{liu2025bridge, xin2025llmcdsr}. X-Cross further explores dynamic knowledge integration among domain-specific language models through low-rank adaptation~\cite{hadad2025xcross}. These studies highlight the value of semantic knowledge and language-based preference modeling for cross-domain recommendation. Different from directly adapting language models or integrating their internal representations, BGAFlow employs LLM-inferred profiles to establish a shared semantic foundation and guide personalized transfer. 

\subsection{Flow Matching}

Flow matching has emerged as an effective framework for learning continuous transformations between probability distributions. It trains time-dependent vector fields along prescribed probability paths without requiring numerical simulation during conditional velocity supervision~\cite{lipman2022flow}. Related formulations, such as rectified flow, further investigate transport trajectories for efficient generation and data transfer~\cite{liu2022flow}. By learning the evolution of intermediate states through a velocity field, flow-based approaches provide a flexible alternative to direct representation mapping.

Recently, flow matching has been introduced into recommender systems for modeling user preferences and generating recommendation representations. FlowCF incorporates behavior-guided priors and discrete flow modeling into collaborative filtering~\cite{liu2025flowcf}. FMRec employs a straight flow trajectory for sequential recommendation, while FlowRec models preference trajectories using personalized behavior-based priors~\cite{liu2025fmrec,li2025preference}. More recent studies investigate additional properties of the learned dynamics. PDFlow introduces a residual vector field to mitigate popularity bias during sequential recommendation, whereas ConFM employs contrastive supervision to preserve user-specific distinctions in flow predictions~\cite{yang2026pdflow,jin2026contrastive}. Although these approaches demonstrate the potential of flow-based recommendation, their primary objectives differ from personalized cross-domain semantic adaptation. 


\begin{figure*}[t]
    \centering
    \includegraphics[width=0.9\columnwidth]{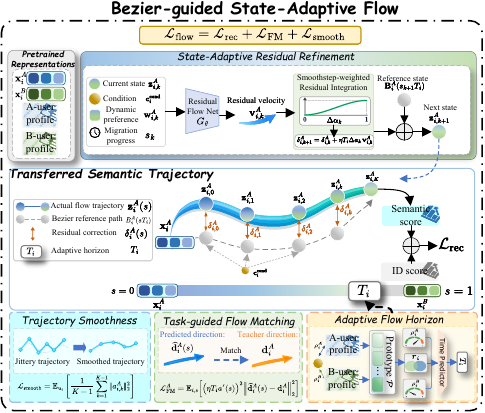}
\caption{Overall architecture of BGAFlow. Unified dual-domain pretraining learns domain-specific collaborative representations and comparable semantic representations. During cross-domain transfer, a profile-conditioned B\'ezier curve provides a geometric reference, while a personalized horizon controls the transfer extent and residual correction scale. A state-adaptive residual flow progressively refines the reference trajectory through dynamic preference feedback and smoothstep-weighted integration. Task-guided flow matching and trajectory smoothness regularization guide the learning process, and the adapted semantic representation is combined with collaborative information for final recommendation.}
\Description{A conditional Bezier reference and a personalized horizon guide a residual velocity network. At each integration step, the current semantic state retrieves prototype feedback and receives a smoothstep-weighted correction. Task-guided flow matching, recommendation loss, and trajectory regularization train the model.}
\label{fig:BGAFlow}
\end{figure*}
\section{Methodology}
\subsection{Problem Formulation}
We study cross-domain sequential recommendation with two domains, denoted as \textit{A} and \textit{B}. Let $\mathcal{U}^A$ and $\mathcal{U}^B$ be the user sets in the two domains, and let $\mathcal{U}^O = \mathcal{U}^A \cap \mathcal{U}^B$ denote the set of overlapping users. The item sets of the two domains are denoted as $\mathcal{V}^A$ and $\mathcal{V}^B$, respectively. Following the common cross-domain recommendation setting, we assume that the two domains have disjoint item sets, i.e., $\mathcal{V}^{A} \cap \mathcal{V}^{B} = \emptyset$.  For each overlapping user $u_i \in \mathcal{U}^O$, we observe two domain-specific interaction sequences, denoted as $\mathcal{H}_i^A$ and $\mathcal{H}_i^B$. Specifically, $\mathcal{H}_i^A = \{v_{i,1}^A, v_{i,2}^A, \ldots, v_{i,n_i^A}^A\}$ is the user's historical interaction sequence in domain \textit{A}, where each item $v_{i,j}^A \in \mathcal{V}^A$. Similarly, $\mathcal{H}_i^B = \{v_{i,1}^B, v_{i,2}^B, \ldots, v_{i,n_i^B}^B\}$ denotes the interaction sequence in domain \textit{B}. Each sequence is ordered according to interaction time.
Our goal is to improve the recommendation performance in one domain by exploiting useful preference information from the other domain. The overall framework is illustrated in Fig.~\ref{fig:BGAFlow}.
\begin{figure*}
    \centering
    \includegraphics[width=0.7\textwidth]{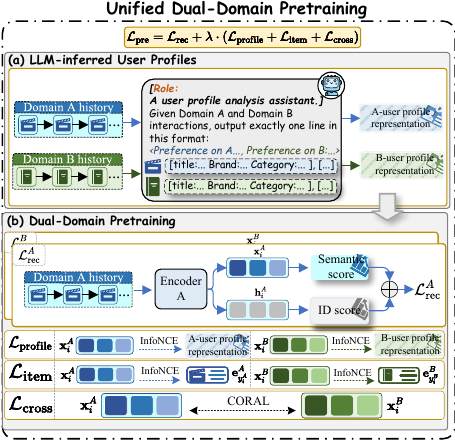}
\caption{Overview of unified dual-domain pretraining. (a) An LLM infers domain-specific user profiles from interaction histories in both domains. (b) A dual-tower architecture learns domain-specific collaborative representations and shared semantic representations for next-item prediction. Profile and item-semantic alignment objectives enrich semantic representations, while CORAL regularization improves cross-domain distribution consistency, providing a comparable semantic space for subsequent transfer.}
\Description{Panel (a) converts interaction histories from two domains into domain-specific user profiles with a language model. Panel (b) combines collaborative and semantic scores for next-item prediction and applies profile alignment, item alignment, and cross-domain covariance regularization during pretraining.}
\label{fig:pretraining}
\end{figure*}
\subsection{System Overview}
We propose BGAFlow for cross-domain sequential recommendation. BGAFlow aims to replace static cross-domain alignment with a personalized and state-adaptive semantic adaptation process. It consists of two main components: Unified Dual-Domain Pretraining and B\'ezier-guided State-Adaptive Flow.

\textbf{1) The Unified Dual-Domain Pretraining.} This component is designed to provide comparable and semantically enriched features for cross-domain alignment. We first use an LLM to infer domain-level user interest profiles from users’ historical interactions in two domains. Then, BGAFlow adopts an ID-text dual-tower encoder, where the ID tower is domain-specific and the text tower is shared across domains. During pretraining, we introduce three auxiliary objectives for the semantic branch. The user-profile alignment objective encourages semantic representations to capture richer user preference semantics from LLM-inferred profiles. The item-semantic alignment objective aligns semantic representations with the ground-truth item semantics, making them more relevant to next-item prediction. In addition, the cross-domain correlation alignment (CORAL) regularization constrains the semantic representation distributions of the two domains, providing a more comparable space for subsequent alignment.

\textbf{2) The B\'ezier-guided State-Adaptive Flow.} Rather than learning to reproduce the analytically known velocity of a prescribed B\'ezier curve, this component uses the curve as a geometric reference and learns task-relevant residual corrections. A profile-conditioned control point specifies the reference path, while a prototype-based horizon predictor determines a user-specific transfer extent. During adaptation, the residual velocity field receives the current representation, transfer progress, condition feature, and dynamically retrieved preference prototypes, allowing subsequent corrections to depend on intermediate states. A training-only task-guided path provides residual-velocity supervision, and a smoothstep schedule controls the progressive application of corrections. The adapted semantic representation is then combined with the collaborative representation for next-item recommendation.

\subsection{Unified Dual-Domain Pretraining}

The flow matching module requires the two endpoint representations to be comparable and located in a coherent feature space. Therefore, BGAFlow performs cross-domain alignment on semantically oriented representations rather than ID-based collaborative representations. As illustrated in Fig.~\ref{fig:pretraining}, unified dual-domain pretraining combines LLM-inferred user profiles with dual-domain representation learning to construct a shared semantic space, providing suitable endpoints for subsequent state-adaptive alignment.

\subsubsection{\textbf{LLM-inferred User Profiles}}\label{sec:user_profiles}

Although semantics-enhanced sequential encoders (e.g.,~\cite{hou2022towards,yuan2023go}) can capture semantic information from user interactions, the learned representations are still heavily influenced by domain-specific semantics. Therefore, we use an LLM~\cite{zhao2023survey, wu2024survey} to infer domain-level user profiles from historical interactions using the prompt template given below. For user $u_i \in \mathcal{U}^O$, let $\mathcal{H}_i^d$ denote the historical interaction sequence in domain $d \in \{A,B\}$. Each interacted item is converted into its available textual metadata, such as title, category, brand, and description. Given the textual histories from both domains, excluding ground-truth items, we prompt the LLM to infer the user's domain-level profiles:
\begin{equation}
    (\bm{c}_i^A, \bm{c}_i^B) = \mathrm{LLM}(\mathrm{Text}(\mathcal{H}_i^A), \mathrm{Text}(\mathcal{H}_i^B), \mathrm{Prompt}),
\end{equation}
where $\bm{c}_i^A$ and $\bm{c}_i^B$ describe the user's profiles in domains \textit{A} and \textit{B}, respectively, and $\mathrm{Prompt}$ denotes the instruction template used to guide the LLM. We use the following prompt, replacing \texttt{[Interaction text]} with the textual histories from the two domains:

\begin{quote}
\setlength{\parskip}{3pt}
\noindent\textbf{Role:} You are a cross-domain user profile analysis assistant.

\noindent\textbf{Task:} Given Domain A and Domain B interactions, output exactly one line in this strict format:

\noindent\texttt{\{Preference on A: ..., Preference on B: ...\}}

\noindent\textbf{Rules:}
\begin{enumerate}[label=(\arabic*),leftmargin=*,nosep,topsep=2pt]
    \item Preference on A must describe only Domain A preferences.
    \item Preference on B must describe only Domain B preferences.
    \item If a domain has clear interactions, do not write `Insufficient evidence.' for that domain.
    \item Do not include brand names, product IDs, URLs, or copied titles.
    \item Be concise and generalize to reusable preference traits. Write a fresh user-specific summary from the evidence instead of reusing stock phrasing.
\end{enumerate}

\noindent\textbf{Example:} \{Preference on A: prefers weather-specific outdoor training gear, Preference on B: prefers casual apparel suited to seasonal daily wear\}

\noindent\textbf{Input:} \texttt{[Interaction text]}
\end{quote}

\subsubsection{\textbf{Dual-domain Pretraining}}

We adopt a Transformer~\cite{vaswani2017attention}-based ID-text dual-tower encoder. The ID tower is domain-specific, while the text tower is shared across domains to produce semantic representations in the same feature space. Taking domain \textit{A} as an example, the collaborative representation is obtained as $\bm{h}_i^A=\mathrm{Encoder}_{\text{ID}}^A(\mathcal{H}_i^A)$. For the text modality, we use a mixture-of-experts (MoE) projection. We denote the item text embedding sequence as $\bm{E}_i^A=(\bm{e}_{i,1}^A,\ldots,\bm{e}_{i,n_i^A}^A)$, and obtain the semantic representation by $\bm{x}_i^A=\mathrm{Encoder}_{\text{text}}(\mathrm{MoE}(\bm{E}_i^A))$. Here, $\bm{h}_i^A$ denotes the ID-based collaborative representation and $\bm{x}_i^A$ denotes the semantic representation from the shared text tower. The representations in domain \textit{B} are obtained analogously.

For next-item prediction, we combine the ID-based score and the semantic score. Let $\bm{E}_\text{id}^A$ and $\bm{E}_\text{text}^A$ denote the ID item embedding table and the projected text item embedding table in domain \textit{A}, respectively. Given the ground-truth next item $y_i^A$, the recommendation loss is defined as:
\begin{equation}\label{eq:ce_loss}
\begin{aligned}
\mathcal{L}^A_\mathrm{rec}
&= \sum_{u_i\in\mathcal{U}^O}\mathrm{CE}\left(\bm{o}_i^A,y_i^A\right),\\
\bm{o}_i^A
&= \bm{h}_i^A(\bm{E}_\text{id}^A)^\top
 + \bm{x}_i^A(\bm{E}_\text{text}^A)^\top,
\end{aligned}
\end{equation}
where the loss for domain \textit{B} is defined in the same way, and $\mathcal{L}_\mathrm{rec}=\mathcal{L}^A_\mathrm{rec}+\mathcal{L}^B_\mathrm{rec}$. 
Besides the recommendation loss, we introduce three auxiliary losses to enhance the semantic branch. 
First, the profile alignment loss injects LLM-inferred high-level preference semantics into the semantic representation. This objective encourages the semantic representation to capture abstract user interests beyond item-level textual patterns. Let $\bm{p}_i^A$ be the embedding of the LLM-inferred profile $\bm{c}_i^A$, and let $g(\cdot)$ be a shared projection function that maps profile embeddings into the semantic space. For a mini-batch $\mathcal{B}$, the profile InfoNCE~\cite{yu2022graph, xie2022contrastive, zhang2025unveiling, qin2024hyperbolic} alignment loss in domain \textit{A} is formulated as:
\begin{equation}
    \mathcal{L}_{\mathrm{profile}}^A=-\sum_{u_i\in\mathcal{B}}\log\frac{\exp\left(\mathrm{sim}\left(\bm{x}_i^A,g(\bm{p}_i^A)\right)\right)}{\sum_{u_j\in\mathcal{B}}\exp\left(\mathrm{sim}\left(\bm{x}_i^A,g(\bm{p}_j^A)\right)\right)}.
\end{equation}
Second, the item-semantic alignment loss makes the semantic representation more relevant to the next-item prediction target. Let $\bm{e}^A_{y_i^A}$ be the projected text embedding of the ground-truth item $y_i^A$. For a mini-batch $\mathcal{B}$, the item-semantic InfoNCE loss is:
\begin{equation}
    \mathcal{L}_{\mathrm{item}}^A=-\sum_{u_i\in\mathcal{B}}\log\frac{\exp\left(\mathrm{sim}\left(\bm{x}_i^A,\bm{e}^A_{y_i^A}\right)\right)}{\sum_{u_j\in\mathcal{B}}\exp\left(\mathrm{sim}\left(\bm{x}_i^A,\bm{e}^A_{y_j^A}\right)\right)}.
\end{equation}
Third, we use CORAL regularization as a light distribution-level constraint to improve the comparability of semantic representations across domains, without forcing paired user representations to be identical. Let $\bm{X}^A \in \mathbb{R}^{|\mathcal{B}| \times D}$ and $\bm{X}^B \in \mathbb{R}^{|\mathcal{B}| \times D}$ be the batch semantic representation matrices of domains \textit{A} and \textit{B}, respectively. Their covariance matrices are denoted as $\mathrm{Cov}(\bm{X}^A)$ and $\mathrm{Cov}(\bm{X}^B)$. The CORAL loss is defined as:
\begin{equation}
    \mathcal{L}_{\mathrm{cross}}=\frac{1}{4D^2}\left\|\mathrm{Cov}\left(\bm{X}^A\right)-\mathrm{Cov}\left(\bm{X}^B\right)\right\|_F^2.
\end{equation}
The two domain-specific alignment losses are combined as $\mathcal{L}_{\mathrm{profile}}=\mathcal{L}_{\mathrm{profile}}^A+\mathcal{L}_{\mathrm{profile}}^B$ and $\mathcal{L}_{\mathrm{item}}=\mathcal{L}_{\mathrm{item}}^A+\mathcal{L}_{\mathrm{item}}^B$. Finally, the total pretraining objective is:
\begin{equation}
    \mathcal{L}_\text{pre} =  \mathcal{L}_\text{rec} + \lambda \cdot(\mathcal{L}_{\mathrm{profile}} +  \mathcal{L}_{\mathrm{item}} +  \mathcal{L}_{\mathrm{cross}}).
\end{equation}

\subsection{B\'ezier-guided State-Adaptive Flow}
After unified dual-domain pretraining, the semantic representations of the two domains lie in a comparable feature space, providing a basis for cross-domain preference transfer. A conditional B\'ezier curve offers a smooth geometric reference that incorporates domain-specific semantics and user-profile conditions. However, once its endpoints and control point are fixed, the local directions of this reference curve are predetermined and cannot respond to changes in the representation state during transfer. We therefore retain the B\'ezier curve as a reference and learn a state-adaptive residual velocity field to progressively refine it. The framework further uses a personalized horizon to control the reference transfer extent and a task-guided objective to supervise residual corrections.

\subsubsection{\textbf{Adaptive Flow Horizon}}
For different users, the amount of useful cross-domain information can vary substantially. We use an adaptive horizon $T_i\in(0,1)$ to determine the effective extent of each user's transfer along the B\'ezier reference. A larger $T_i$ traverses a longer portion of the reference path, whereas a smaller $T_i$ leads to more conservative adaptation. We estimate this horizon from prototype-level preference overlap and condition-related geometric similarities.

To characterize latent interest overlap, we perform spherical clustering over the item semantic embeddings from both domains to obtain a shared prototype set $\mathcal{P}=\{\bm{\mu}_1,\ldots,\bm{\mu}_{n_p}\}$, where each $\bm{\mu}_j$ represents a latent semantic interest direction. Let $\bm{p}_i^d$ denote the embedding of the LLM-inferred profile introduced in the pretraining stage, and let $\bm{q}_i^d=g(\bm{p}_i^d)$ be its representation in the shared semantic space for $d\in\{A,B\}$. The prototype preference distribution is defined as
\begin{equation}
    \pi_i^d(j)=\frac{\exp\!\left(\mathrm{sim}(\bm{q}_i^d,\bm{\mu}_j)\right)}{\sum_{l=1}^{n_p}\exp\!\left(\mathrm{sim}(\bm{q}_i^d,\bm{\mu}_l)\right)}.
\end{equation}
The discrepancy between the domain-specific distributions is measured by Jensen--Shannon (JS) divergence:
\begin{equation}
    \mathrm{JS}(\pi_i^A,\pi_i^B)
    =\frac{1}{2}\mathrm{KL}(\pi_i^A\|\bm{m}_i)
    +\frac{1}{2}\mathrm{KL}(\pi_i^B\|\bm{m}_i),
\end{equation}
where $\bm{m}_i=(\pi_i^A+\pi_i^B)/2$. The prototype-level preference overlap is
\begin{equation}
    r_i=1-\frac{\mathrm{JS}(\pi_i^A,\pi_i^B)}{\log 2},\qquad r_i\in[0,1].
\end{equation}

We then construct a profile-conditioned feature that also serves as the B\'ezier control point:
\begin{equation}
    \bm{c}_i^{\mathrm{cond}}=\phi_{\mathrm{cond}}\!\left([\bm{q}_i^A;\bm{q}_i^B]\right),
\end{equation}
where $\phi_{\mathrm{cond}}$ is a learnable multilayer perceptron (MLP). Its geometric relations with the two profile representations are measured by
\begin{equation}
    \rho_i^A=\mathrm{sim}(\bm{c}_i^{\mathrm{cond}},\bm{q}_i^A),\qquad
    \rho_i^B=\mathrm{sim}(\bm{c}_i^{\mathrm{cond}},\bm{q}_i^B).
\end{equation}
The user-specific horizon is predicted as
\begin{equation}\label{eq:adaptive_horizon}
    T_i=\mathrm{Sigmoid}\!\left(\phi_{\mathrm{hori}}(r_i,\rho_i^A,\rho_i^B)\right),
\end{equation}
where $\phi_{\mathrm{hori}}$ is a learnable MLP. In the proposed residual dynamics, $T_i$ specifies the terminal point of the B\'ezier reference and scales the allowable correction, while the numerical integration is carried out over a normalized progress variable $s\in[0,1]$. Importantly, the same prototype set $\mathcal{P}$ is reused to provide state-dependent preference feedback during the subsequent flow, without introducing an additional prototype bank.

\subsubsection{\textbf{B\'ezier-guided State-Adaptive Dynamics}}
We describe the transfer toward the paired-domain semantic reference when making recommendations in domain \textit{A}. The reverse direction is formulated analogously. Given the pretrained semantic representations $\bm{x}_i^A$ and $\bm{x}_i^B$, the condition-guided quadratic B\'ezier reference is defined by
\begin{equation}\label{eq:bezier_path}
    \bm{B}_i^A(t)=(1-t)^2\bm{x}_i^A
    +2(1-t)t\bm{c}_i^{\mathrm{cond}}+t^2\bm{x}_i^B,
    \quad t\in[0,1].
\end{equation}
Its tangent is available in closed form:
\begin{equation}\label{eq:bezier_velocity}
\begin{aligned}
    \bm{\xi}_i^A(t)
    =\frac{d\bm{B}_i^A(t)}{dt}=2(1-t)(\bm{c}_i^{\mathrm{cond}}-\bm{x}_i^A)
    +2t(\bm{x}_i^B-\bm{c}_i^{\mathrm{cond}}).
\end{aligned}
\end{equation}
Unlike direct B\'ezier interpolation or velocity regression to this known tangent, our method uses $\bm{B}_i^A$ as a geometric reference while learning a task-adaptive deviation from it. In particular, we parameterize the actual representation at normalized transfer progress $s$ as
\begin{equation}\label{eq:state_decomposition}
    \bm{z}_i^A(s)=\bm{B}_i^A(sT_i)+\bm{\delta}_i^A(s),
    \qquad \bm{\delta}_i^A(0)=\bm{0},
\end{equation}
where $\bm{\delta}_i^A(s)$ is the accumulated residual correction and $\bm{z}_i^A(s)$ is the actual intermediate semantic representation. This decomposition allows the learned flow to refine a structured reference without numerically reconstructing the analytically known B\'ezier component.

To enable state-aware adaptation, we query the shared preference prototypes using the \emph{current} intermediate representation. Specifically, we compute
\begin{equation}\label{eq:dynamic_preference}
\begin{aligned}
    \alpha_{ij}^A(s)
    &=\frac{\exp\!\left(\mathrm{sim}(\bm{z}_i^A(s),\bm{\mu}_j)\right)}{\sum_{l=1}^{n_p}\exp\!\left(\mathrm{sim}(\bm{z}_i^A(s),\bm{\mu}_l)\right)},\\
    \bm{w}_i^A(s)&=\sum_{j=1}^{n_p}\alpha_{ij}^A(s)\bm{\mu}_j.
\end{aligned}
\end{equation}
Because $\bm{w}_i^A(s)$ is recomputed from $\bm{z}_i^A(s)$, subsequent velocity predictions can respond to the intermediate transfer state, rather than relying only on fixed profile conditions. We denote this prototype-feedback operation by $\mathcal{A}(\bm{z};\mathcal{P})$ in what follows.

The residual dynamics are parameterized as
\begin{equation}\label{eq:residual_dynamics}
    \frac{d\bm{\delta}_i^A(s)}{ds}
    =\eta T_i a'(s)\,
    \bm{G}_{\theta}\!\left(
    \bm{z}_i^A(s),s,\bm{c}_i^{\mathrm{cond}},\bm{w}_i^A(s)
    \right),
\end{equation}
where $\bm{G}_{\theta}$ is a learnable residual velocity network, $\eta>0$ controls the overall correction scale, and the normalized time $s$ is provided through a time embedding. To progressively activate and attenuate residual corrections, we adopt the cubic smoothstep schedule
\begin{equation}\label{eq:smoothstep}
    a(s)=3s^2-2s^3,\qquad a'(s)=6s(1-s).
\end{equation}
The schedule satisfies $a(0)=0$, $a(1)=1$, and $a'(0)=a'(1)=0$, so the additional residual velocity is smoothly gated at the two boundaries. This scheduling does not assume that the B\'ezier reference is nonsmooth. Instead, it controls how the learned correction is introduced along an already smooth reference trajectory.

\subsubsection{\textbf{Task-guided Flow Matching}}
Directly supervising $\bm{G}_{\theta}$ with the known B\'ezier tangent would require the residual network to reproduce information already available from Eq.~(\ref{eq:bezier_velocity}). We instead construct a task-guided conditional path from the recommendation objective, thereby providing a supervision signal for corrections beyond the geometric reference.

Let $\overline{\bm{x}}_i^A=\bm{B}_i^A(T_i)$ denote the reference terminal representation without residual adaptation. Using the same recommendation head as the pretraining stage, its per-user reference loss is
\begin{equation}\label{eq:reference_rec_loss}
    \ell_{i,\mathrm{ref}}^A
    =\mathrm{CE}\!\left(
    \bm{h}_i^A(\bm{E}_{\mathrm{id}}^A)^\top
    +\overline{\bm{x}}_i^A(\bm{E}_{\mathrm{text}}^A)^\top,
    y_i^A\right).
\end{equation}
We obtain one task-guided correction direction for user $u_i$ by taking the negative gradient of this reference loss with respect to the reference representation:
\begin{equation}\label{eq:task_direction}
    \bm{d}_i^A
    =-\operatorname{sg}\!\left(
    \operatorname{clip}_{\gamma}\!\left(
    \nabla_{\overline{\bm{x}}_i^A}\ell_{i,\mathrm{ref}}^A
    \right)\right),
\end{equation}
where $\operatorname{sg}(\cdot)$ stops gradients through the teacher direction and $\operatorname{clip}_{\gamma}(\cdot)$ bounds its norm for stable supervision. Importantly, $\bm{d}_i^A$ is calculated \emph{once} for each training instance at the reference terminal state and remains fixed over its entire teacher trajectory, and no intermediate-state task gradients are computed.

We construct the task-guided conditional path as
\begin{equation}\label{eq:task_guided_path}
    \widetilde{\bm{z}}_i^A(s)
    =\bm{B}_i^A(sT_i)+\eta T_i a(s)\bm{d}_i^A.
\end{equation}
The corresponding teacher residual velocity is analytically available:
\begin{equation}\label{eq:teacher_residual_velocity}
    \frac{d}{ds}\!\left(
    \widetilde{\bm{z}}_i^A(s)-\bm{B}_i^A(sT_i)
    \right)=\eta T_i a'(s)\bm{d}_i^A.
\end{equation}
For a sampled $s\sim\mathcal{U}(0,1)$, we calculate the preference feedback on the teacher state as $\widetilde{\bm{w}}_i^A(s)=\mathcal{A}(\widetilde{\bm{z}}_i^A(s);\mathcal{P})$ and predict its correction direction by
\begin{equation}\label{eq:predicted_teacher_direction}
\begin{aligned}
    \widehat{\bm{d}}_i^A(s)
    =\bm{G}_{\theta}\!\bigl(\widetilde{\bm{z}}_i^A(s),s,\bm{c}_i^{\mathrm{cond}},\widetilde{\bm{w}}_i^A(s)\bigr).
\end{aligned}
\end{equation}
We supervise the corresponding residual velocity using the task-guided flow matching objective:
\begin{equation}\label{eq:task_fm_loss}
    \mathcal{L}_{\mathrm{FM}}^A
    =\mathbb{E}_{i,s}\!\left[
    \big(\eta T_i a'(s)\big)^2
    \left\|\widehat{\bm{d}}_i^A(s)-\bm{d}_i^A\right\|_2^2
    \right].
\end{equation}
Unlike matching the B\'ezier tangent itself, this objective teaches the network to predict task-relevant residual corrections from observable states and preference conditions. During inference, neither the ground-truth next item nor the task-gradient teacher is available. The learned velocity network is evaluated using only the current representation, time, condition feature, and prototype feedback. Although the teacher direction is fixed per training instance, the student velocity network remains state-conditioned and can respond to its evolving rollout states.

\subsubsection{\textbf{Discrete Residual Integration and Recommendation}}
We numerically integrate only the residual dynamics, while the B\'ezier reference is evaluated directly through Eq.~(\ref{eq:bezier_path}). For $K$ integration steps, let $s_k=k/K$ and define the smoothstep increment
\begin{equation}\label{eq:delta_a}
    \Delta a_k=a(s_{k+1})-a(s_k),\qquad k=0,\ldots,K-1.
\end{equation}
Starting from $\bm{\delta}_{i,0}^A=\bm{0}$, the discrete update is
\begin{equation}\label{eq:residual_rollout}
\begin{aligned}
    \bm{z}_{i,k}^A
    &=\bm{B}_i^A(s_kT_i)+\bm{\delta}_{i,k}^A,\\
    \bm{w}_{i,k}^A
    &=\mathcal{A}(\bm{z}_{i,k}^A;\mathcal{P}),\\
    \bm{v}_{i,k}^A
    &=\bm{G}_{\theta}(\bm{z}_{i,k}^A,s_k,
    \bm{c}_i^{\mathrm{cond}},\bm{w}_{i,k}^A),\\
    \bm{\delta}_{i,k+1}^A
    &=\bm{\delta}_{i,k}^A
    +\eta T_i\Delta a_k\,\bm{v}_{i,k}^A.
\end{aligned}
\end{equation}
Instead of approximating the schedule increment by $a'(s_k)/K$, this update uses its exact difference over each interval, with $\sum_{k=0}^{K-1}\Delta a_k=a(1)-a(0)=1$. In particular, if the predicted residual direction is constant across steps, the cumulative schedule weight is exact for any $K$. For state-dependent velocities, Eq.~(\ref{eq:residual_rollout}) remains a numerical approximation to the residual dynamics. The final adapted representation is
\begin{equation}\label{eq:adapted_representation}
    \widetilde{\bm{x}}_i^A
    =\bm{z}_{i,K}^A
    =\bm{B}_i^A(T_i)+\bm{\delta}_{i,K}^A.
\end{equation}
It is combined with the original collaborative representation for next-item scoring:
\begin{equation}\label{eq:adapted_score}
    \bm{o}_i^A
    =\bm{h}_i^A(\bm{E}_{\mathrm{id}}^A)^\top
    +\widetilde{\bm{x}}_i^A(\bm{E}_{\mathrm{text}}^A)^\top,
\end{equation}
with $\mathcal{L}_{\mathrm{rec}}^A=\sum_{u_i\in\mathcal{U}^O}\mathrm{CE}(\bm{o}_i^A,y_i^A)$.

To encourage gradual evolution of the actual transferred representation, we regularize the discrete acceleration of the complete transfer trajectory. The rollout provides $K+1$ states $\{\bm{z}_{i,k}^A\}_{k=0}^{K}$, where each state includes both the B\'ezier reference and the accumulated residual as defined in Eq.~(\ref{eq:residual_rollout}). Since $s_k=k/K$, the second-order finite difference with respect to the normalized progress $s$ is
\begin{equation}\label{eq:trajectory_acceleration}
\begin{aligned}
    \bm{a}_{i,k}^A=K^2\bigl(&\bm{z}_{i,k+1}^A
    -2\bm{z}_{i,k}^A+\bm{z}_{i,k-1}^A\bigr),
    \quad k=1,\ldots,K-1.
\end{aligned}
\end{equation}
The trajectory smoothness objective is defined as
\begin{equation}\label{eq:trajectory_smoothness}
    \mathcal{L}_{\mathrm{smooth}}^A
    =\mathbb{E}_{u_i}\!\left[
    \frac{1}{K-1}\sum_{k=1}^{K-1}
    \left\|\bm{a}_{i,k}^A\right\|_2^2
    \right],\qquad K\geq 2.
\end{equation}
Unlike a regularizer on residual velocities alone, this objective penalizes abrupt changes in the evolution of the \emph{complete} semantic state. Consequently, it accounts for curvature induced by both the B\'ezier reference and the residual correction, including the temporal modulation introduced by the smoothstep schedule. Its weight is therefore controlled to balance smooth transfer dynamics against preservation of useful reference-path geometry and recommendation accuracy.

The overall objective of the flow-based recommendation stage is
\begin{equation}\label{eq:flow_loss}
    \mathcal{L}_{\mathrm{flow}}^A
    =\mathcal{L}_{\mathrm{rec}}^A
    +\mathcal{L}_{\mathrm{FM}}^A
    +\mathcal{L}_{\mathrm{smooth}}^A.
\end{equation}
The reverse recommendation direction is implemented analogously by exchanging domains \textit{A} and \textit{B} and constructing its own reference path and task-guided correction direction.


\subsubsection{\textbf{Theoretical Analysis}}
\label{sec:theoretical_analysis}

The proposed B\'ezier-guided state-adaptive flow separates the geometric
reference from the learned residual correction. This decomposition provides
an interpretable basis for analyzing how the personalized horizon regulates
the extent of cross-domain adaptation and how the smoothstep schedule
controls the accumulation of residual updates. We consider the transfer
for recommendation in domain \textit{A}. The same analysis applies to the reverse direction.

We first examine the deviation of the transferred representation from its
B\'ezier reference. As shown in Eq.~(\ref{eq:residual_rollout}), each
residual update is weighted by the nonnegative increment $\Delta a_k$.
Unrolling the discrete dynamics from $\bm{\delta}_{i,0}^A=\bm{0}$ gives
\begin{equation}\label{eq:accumulated_residual}
    \bm{\delta}_{i,k}^A
    =\eta T_i\sum_{j=0}^{k-1}\Delta a_j\bm{v}_{i,j}^A.
\end{equation}
This expression makes explicit that the personalized horizon controls
the scale of the correction, while the schedule determines how the
predicted directions are accumulated. Suppose that the predicted
residual velocities satisfy $\|\bm{v}_{i,k}^A\|_2\leq M$ along the
rollout. Since $\sum_{j=0}^{k-1}\Delta a_j=a(s_k)$, the triangle
inequality bounds the accumulated residual by $\eta T_i M a(s_k)$.
Furthermore, let
$M_{B,i}=\sup_{t\in[0,1]}\|\frac{d}{dt}\bm{B}_i^A(t)\|_2$
denote the maximum speed along the B\'ezier reference. Using
$\bm{B}_i^A(0)=\bm{x}_i^A$, we obtain
\begin{equation}\label{eq:controlled_transfer_bound}
\begin{aligned}
    \|\bm{z}_{i,k}^A-\bm{B}_i^A(s_kT_i)\|_2
    &\leq \eta T_i M a(s_k),\\
    \|\bm{z}_{i,K}^A-\bm{x}_i^A\|_2
    &\leq T_i(M_{B,i}+\eta M).
\end{aligned}
\end{equation}
The first inequality bounds the learned correction relative to the
geometric reference, whereas the second bounds the total adaptation
relative to the source-domain representation. Both bounds scale with
$T_i$, indicating that the personalized horizon regulates not only how
far the reference path is traversed but also the maximum magnitude of
the residual refinement. Thus, a shorter horizon imposes a more
conservative upper bound on representation changes. These bounds hold
under the stated velocity condition and do not, by themselves, guarantee
that negative transfer is avoided.

The temporal schedule also provides a useful consistency property for
discrete integration. The smoothstep function
$a(s)=3s^2-2s^3$ increases from $a(0)=0$ to $a(1)=1$.
Consequently, its exact increments telescope, independently of the
number of integration steps. In particular, when the residual velocity
remains constant at $\bm{v}$ throughout the rollout, we have
\begin{equation}\label{eq:exact_schedule_accumulation}
    \sum_{k=0}^{K-1}\Delta a_k=1,
    \qquad
    \bm{\delta}_{i,K}^A=\eta T_i\bm{v}.
\end{equation}
This result holds for any $K\geq1$: changing the number of steps does
not alter the total correction merely because the smoothstep schedule
has been discretized. By comparison, a left-endpoint Euler update based
on $a'(s_k)/K$ accumulates only a factor of $1-1/K^2$ for the same
constant velocity. The exact-increment formulation therefore removes
this schedule-induced scaling error. When the predicted velocity varies
with the evolving state, the rollout is still a numerical approximation
to the residual dynamics, rather than an exact solution.

Finally, smoothstep satisfies $a'(0)=a'(1)=0$. Provided that the
velocity field remains bounded near the endpoints, the continuous
residual dynamics in Eq.~(\ref{eq:residual_dynamics}) therefore satisfy
\begin{equation}\label{eq:boundary_residual_velocity}
    \left.\frac{d\bm{\delta}_i^A(s)}{ds}\right|_{s=0}
    =\left.\frac{d\bm{\delta}_i^A(s)}{ds}\right|_{s=1}=\bm{0}.
\end{equation}
This zero-velocity boundary property concerns the
additional correction, not the complete transfer trajectory, whose
motion also includes the B\'ezier reference. The full-trajectory
smoothness regularizer further constrains the discrete acceleration of
the combined states $\bm{z}_{i,k}^A$, complementing the scheduling
property by encouraging gradual evolution of the actual transferred
representation. Together, these analytical properties explain the
respective roles of geometric guidance, personalized adaptation, and
temporal scheduling without assuming an unconditional improvement in
recommendation accuracy.

\begin{table}
\setlength{\abovecaptionskip}{0.0cm}
\setlength{\belowcaptionskip}{0.0cm} 
\centering
\caption{Statistics of datasets.}
\label{tab:datasets}
\resizebox{0.65\columnwidth}{!}{
\begin{tabular}{lccccc}
\toprule
\textbf{Datasets} & \textbf{\#(Users)} & \textbf{\#(Items)} & \textbf{\#(Actions)} & \textbf{Avg.} & \textbf{Sparsity} \\
                  &                    &                    &                       & \textbf{SeqLen.} &                  \\
\midrule
Movie & \multirow{2}{*}{19,577} & 34,384 & 158,814 & 8.11 & 99.976\% \\
Book  &                        & 49,113  & 185,406  & 9.47  & 99.981\% \\
\midrule
Sport & \multirow{2}{*}{23,330} & 17,295 & 196,760 & 8.43 & 99.951\% \\
Cloth  &                        & 37,175 & 198,454 & 8.51 & 99.977\% \\
\bottomrule
\end{tabular}
}
\end{table}

\section{Experiments}
\subsection{Experimental Settings}
\subsubsection{Datasets and Evaluation Protocol}
Following previous work on cross-domain recommendation~\cite{xin2025llmcdsr, xu2025multi}, we construct two cross-domain sequential recommendation datasets from the widely used public Amazon Review dataset\footnote{\url{https://nijianmo.github.io/amazon/index.html}}: \textbf{Movie-Book} and \textbf{Sport-Cloth}. Since BGAFlow relies on item textual information, we first remove items without available text metadata. We treat user ratings as implicit feedback and retain only interactions with ratings no less than 4. For each domain pair, we select users who have interactions in both domains to form the cross-domain dataset. We keep each user’s interactions within the most recent year and sort them chronologically. Following standard preprocessing procedures, we iteratively filter out users and items with fewer than 5 interactions, and further require each user to have more than 2 interactions in both domains. The statistics of the processed datasets are reported in Table~\ref{tab:datasets}.

For evaluation, we take the last interacted item of each user in each domain as the ground-truth item. During validation and testing, for each ground-truth item, we randomly sample 999 negative items from the corresponding domain and combine them with the ground-truth item to form the candidate set. We evaluate recommendation performance using hit ratio (HR@$N$) and normalized discounted cumulative gain (NDCG@$N$), where the ranking cutoff $N$ is set to 5 and 10.

\subsubsection{Baselines}
To comprehensively evaluate the effectiveness of BGAFlow, we compare BGAFlow with the following types of methods:

\textbf{Conventional single-domain methods:}
\begin{itemize}[label=--, leftmargin=*]
    \item \textbf{SASRec}~\cite{kang2018self} is a classic single-domain sequential recommendation model that uses self-attention to capture users’ sequential preference patterns from historical interactions.
    \item \textbf{FEARec}~\cite{du2023frequency} enhances sequential recommendation by modeling user behavior patterns in both time and frequency domains with hybrid attention and frequency-aware regularization.
    \item \textbf{UniSRec}~\cite{hou2022towards} learns transferable sequence representations by using item textual descriptions and contrastive pre-training across domains or platforms.
\end{itemize}

\textbf{Conventional cross-domain methods:}
\begin{itemize}[label=--, leftmargin=*]
    \item \textbf{$\mathbf{\pi}$-Net}~\cite{ma2019pi} performs shared-account cross-domain sequential recommendation by using parallel networks to exchange useful information between two domains.
    \item \textbf{C$^2$DSR}~\cite{cao2022contrastive} captures both intra-domain and inter-domain sequential dependencies and uses contrastive learning to enhance cross-domain user preference representations.
    \item \textbf{Tri-CDR}~\cite{ma2024triple} jointly models source-domain, target-domain, and mixed-domain behavior sequences to capture triple correlations for cross-domain recommendation.
    \item \textbf{RecGURU}~\cite{li2022recguru} learns generalized user representations across domains with a self-attentive autoencoder and adversarial domain discriminator, aiming to transfer domain-general user preferences for cross-domain recommendation.
    \item \textbf{SyNCRec}~\cite{park2024pacer} estimates domain-level negative transfer and adaptively coordinates single-domain and cross-domain sequential recommendation through cooperative learning.
    \item \textbf{ABXI}~\cite{bian2025abxi} uses domain LoRA and invariant LoRA to extract domain-invariant interests from cross-domain sequences and adapt them to task-specific domain recommendation.
\end{itemize}

\textbf{Diffusion-based methods:}
\begin{itemize}[label=--, leftmargin=*]
    \item \textbf{DiffuRec}~\cite{li2023diffurec} introduces diffusion modeling into sequential recommendation by generating target item representations from users’ historical behaviors.
    \item \textbf{DiffCDR}~\cite{xuan2024diffusion} applies diffusion probabilistic models as a cross-domain mapping module to generate target-domain user embeddings from auxiliary-domain information.
    \item \textbf{CD-CDR}~\cite{li2025cd} applies conditional diffusion to generate target-domain item representations for cross-domain recommendation under a unified distribution paradigm.
    \item \textbf{D2TCDR}~\cite{zhou2026d2tcdr} learns domain-invariant preference representations through disentanglement and transfers them with a guided diffusion model for cross-domain recommendation.
\end{itemize}

\textbf{LLM-based methods:}
\begin{itemize}[label=--, leftmargin=*]
    \item \textbf{LLM-ESR}~\cite{liu2024llm} enhances sequential recommendation with semantic embeddings derived from LLMs, especially targeting long-tail user and item issues.
    \item \textbf{LeCDSR}~\cite{wang2025lecdsr} leverages LLM-generated semantic embeddings and cross-domain user profiles to enhance preference transfer in cross-domain sequential recommendation.
    \item \textbf{LLM4CDSR}~\cite{liu2025bridge} uses LLM-based item representation and hierarchical user profiling to bridge domains and model cross-domain sequential preferences.
\end{itemize}
For fair comparison, we adopt the same reasoning model and embedding model as BGAFlow for all LLM-based methods.
\subsubsection{Implementation Details}
We implement BGAFlow with PyTorch and conduct all experiments on an NVIDIA RTX 5090 GPU. We use Qwen3.5-4B\footnote{\url{https://huggingface.co/Qwen/Qwen3.5-4B}} to infer user profiles from historical interactions, and employ Qwen3-Embedding-4B\footnote{\url{https://huggingface.co/Qwen/Qwen3-Embedding-4B}} to encode both user profile texts and item texts into dense embeddings. The output embedding dimension is set to 128.
For BGAFlow, the hidden dimension is set to 128. We train the model with the Adam optimizer~\cite{kingma2014adam}, where the learning rate is set to $1 \times 10^{-4}$ and the batch size is set to 128. In the \textbf{Unified Dual-Domain Pretraining} stage, the Transformer-based sequence encoder contains 2 layers with 1 attention head. The number of experts in the MoE projection layer for text representations is set to 4. The weights of the auxiliary pretraining losses are set to 0.1. We adopt early stopping with a patience of 10, using the average HR@10 over domains A and B on the validation set as the selection criterion. In the \textbf{B\'ezier-guided State-Adaptive Flow} stage, the parameters of the pretrained encoders are kept trainable and further optimized with the flow-based recommendation objective. The number of latent prototypes is set to 128, and we perform $K=16$ rollout steps over normalized time $s\in[0,1]$. With the cubic schedule $a(s)=3s^2-2s^3$, each residual update uses the exact schedule increment
$\Delta a_k=a((k+1)/K)-a(k/K)$, while evaluating the state-dependent velocity at the current state. We also apply early stopping in this stage and save the model with the best validation performance on each dataset.
\begin{table*}
\setlength\tabcolsep{1.5pt}
\renewcommand\arraystretch{1.08}
\centering
\caption{Overall performance comparison on Movie-Book and Sport-Cloth datasets. The best results are indicated in bold, and the second-best results are underlined. Improv. denotes the relative improvement over the strongest baseline in each column. Improvements over baselines are statistically significant with $p < 0.001$.}
\label{tab:compare}
\vspace{-3mm}
\resizebox{\textwidth}{!}{
\begin{tabular}{l*{8}{c}@{\hskip 6pt}*{8}{c}}
\toprule
\multirow{4}{*}{Method} &
  \multicolumn{8}{c}{\textbf{Movie-Book}} &
  \multicolumn{8}{c}{\textbf{Sport-Cloth}} \\
\cmidrule(lr){2-9}\cmidrule(lr){10-17}
& \multicolumn{4}{c}{\textbf{Movie}} &
  \multicolumn{4}{c}{\textbf{Book}} &
  \multicolumn{4}{c}{\textbf{Sport}} &
  \multicolumn{4}{c}{\textbf{Cloth}} \\
\cmidrule(lr){2-5}\cmidrule(lr){6-9}\cmidrule(lr){10-13}\cmidrule(lr){14-17}
& \multicolumn{2}{c}{HR} & \multicolumn{2}{c}{NDCG} &
  \multicolumn{2}{c}{HR} & \multicolumn{2}{c}{NDCG} &
  \multicolumn{2}{c}{HR} & \multicolumn{2}{c}{NDCG} &
  \multicolumn{2}{c}{HR} & \multicolumn{2}{c}{NDCG} \\
\cmidrule(lr){2-3}\cmidrule(lr){4-5}\cmidrule(lr){6-7}\cmidrule(lr){8-9}\cmidrule(lr){10-11}\cmidrule(lr){12-13}\cmidrule(lr){14-15}\cmidrule(lr){16-17}
& H@5 & H@10 & N@5 & N@10 & H@5 & H@10 & N@5 & N@10 & H@5 & H@10 & N@5 & N@10 & H@5 & H@10 & N@5 & N@10 \\
\midrule
SASRec &
  0.4136 &
  0.4715 &
  0.3424 &
  0.3612 &
  0.2150 &
  0.2510 &
  0.1870 &
  0.1985 &
  0.1079 &
  0.1328 &
  0.0892 &
  0.0972 &
  0.1161 &
  0.1353 &
  0.1028 &
  0.1089 \\
FEARec &
  0.4203 &
  0.4782 &
  0.3535 &
  0.3720 &
  0.2258 &
  0.2626 &
  0.1885 &
  0.2002 &
  0.1052 &
  0.1355 &
  0.0943 &
  0.1040 &
  0.1158 &
  0.1295 &
  0.1030 &
  0.1074 \\
UniSRec &
  0.3885 &
  0.4554 &
  0.3223 &
  0.3439 &
  0.2298 &
  0.2767 &
  0.1924 &
  0.2076 &
  0.1130 &
  0.1348 &
  0.0851 &
  0.0921 &
  0.1238 &
  0.1415 &
  0.1038 &
  0.1095 \\
\addlinespace[1pt]
\midrule
$\mathbf{\pi}$-Net &
  0.3562 &
  0.3991 &
  0.2833 &
  0.2970 &
  0.2067 &
  0.2334 &
  0.1832 &
  0.1917 &
  0.0953 &
  0.1189 &
  0.0811 &
  0.0886 &
  0.1131 &
  0.1334 &
  0.1033 &
  0.1098 \\
C$^2$DSR &
  0.3772 &
  0.4101 &
  0.2977 &
  0.3082 &
  0.2286 &
  0.2633 &
  0.1938 &
  0.2049 &
  0.1090 &
  0.1305 &
  0.0912 &
  0.0989 &
  0.1096 &
  0.1298 &
  0.0979 &
  0.1043 \\
Tri-CDR &
  0.4607 &
  0.5215 &
  0.2660 &
  0.2854 &
  0.2749 &
  0.3239 &
  0.2247 &
  0.2420 &
  0.1499 &
  0.1913 &
  0.1360 &
  0.1492 &
  0.1583 &
  0.1833 &
  0.1193 &
  0.1273 \\
RecGURU &
  0.3790 &
  0.4123 &
  0.2954 &
  0.3060 &
  0.1985 &
  0.2114 &
  0.1782 &
  0.1821 &
  0.0969 &
  0.1213 &
  0.0833 &
  0.0911 &
  0.1052 &
  0.1234 &
  0.0971 &
  0.1029 \\
SyNCRec &
  0.4624 &
  {\ul 0.5324} &
  0.3779 &
  0.3994 &
  {\ul 0.3061} &
  0.3596 &
  0.2574 &
  0.2745 &
  0.1592 &
  0.1969 &
  0.1366 &
  0.1486 &
  0.1659 &
  0.2013 &
  0.1242 &
  0.1355 \\
ABXI &
  0.4479 &
  0.4799 &
  0.3432 &
  0.3534 &
  0.3056 &
  {\ul 0.3722} &
  0.2545 &
  0.2757 &
  0.1520 &
  0.1931 &
  {\ul 0.1401} &
  0.1532 &
  {\ul 0.1679} &
  0.2062 &
  0.1253 &
  0.1375 \\
\addlinespace[1pt]
\midrule
DiffuRec &
  0.3966 &
  0.4675 &
  0.3400 &
  0.3626 &
  0.2051 &
  0.2488 &
  0.1813 &
  0.1952 &
  0.1112 &
  0.1396 &
  0.0939 &
  0.1023 &
  0.1281 &
  0.1397 &
  0.1057 &
  0.1094 \\
DiffCDR &
  0.3915 &
  0.4269 &
  0.2834 &
  0.2947 &
  0.2557 &
  0.2923 &
  0.1898 &
  0.2015 &
  0.1258 &
  0.1349 &
  0.0926 &
  0.0955 &
  0.1396 &
  0.1521 &
  0.1083 &
  0.1123 \\
CD-CDR &
  0.4252 &
  0.4583 &
  0.3217 &
  0.3323 &
  0.2810 &
  0.3234 &
  0.2181 &
  0.2316 &
  0.1401 &
  0.1818 &
  0.1137 &
  0.1270 &
  0.1535 &
  0.1805 &
  0.1201 &
  0.1287 \\
D2TCDR &
  0.4383 &
  0.4792 &
  0.3359 &
  0.3493 &
  0.3034 &
  0.3629 &
  0.2293 &
  0.2484 &
  0.1642 &
  0.2128 &
  0.1343 &
  0.1498 &
  0.1596 &
  0.1883 &
  0.1267 &
  0.1359 \\
\addlinespace[1pt]
\midrule
LLM-ESR &
  0.4323 &
  0.5063 &
  0.3472 &
  0.3708 &
  0.2841 &
  0.3260 &
  0.2252 &
  0.2386 &
  0.1571 &
  0.1930 &
  0.1267 &
  0.1382 &
  0.1432 &
  0.1981 &
  0.1212 &
  0.1387 \\
LeCDSR &
  0.4459 &
  0.4933 &
  0.3676 &
  0.3827 &
  0.2894 &
  0.3259 &
  0.2348 &
  0.2464 &
  0.1393 &
  0.1982 &
  0.1155 &
  0.1343 &
  0.1631 &
  0.2053 &
  0.1264 &
  0.1399 \\
LLM4CDSR &
  {\ul 0.4633} &
  0.5319 &
  {\ul 0.3812} &
  {\ul 0.4031} &
  0.3044 &
  0.3717 &
  {\ul 0.2603} &
  {\ul 0.2818} &
  {\ul 0.1786} &
  {\ul 0.2316} &
  0.1392 &
  {\ul 0.1561} &
  0.1675 &
  {\ul 0.2143} &
  {\ul 0.1329} &
  {\ul 0.1478} \\
\addlinespace[1pt]
\midrule
\rowcolor{gray!18} \textbf{BGAFlow} &
  \textbf{0.4880} &
  \textbf{0.5532} &
  \textbf{0.4123} &
  \textbf{0.4333} &
  \textbf{0.3268} &
  \textbf{0.4002} &
  \textbf{0.2713} &
  \textbf{0.2948} &
  \textbf{0.2023} &
  \textbf{0.2701} &
  \textbf{0.1581} &
  \textbf{0.1800} &
  \textbf{0.1921} &
  \textbf{0.2460} &
  \textbf{0.1535} &
  \textbf{0.1708} \\
\textit{Improv. (\%)} &
  \textit{+5.33} &
  \textit{+3.91} &
  \textit{+8.16} &
  \textit{+7.49} &
  \textit{+6.76} &
  \textit{+7.52} &
  \textit{+4.23} &
  \textit{+4.61} &
  \textit{+13.27} &
  \textit{+16.62} &
  \textit{+12.85} &
  \textit{+15.31} &
  \textit{+14.41} &
  \textit{+14.79} &
  \textit{+15.50} &
  \textit{+15.56} \\
\bottomrule
\end{tabular}
}
\label{tab:overall}
\end{table*}

\subsection{Performance Comparisons}

Table~\ref{tab:overall} reports the overall recommendation performance on the Movie-Book and Sport-Cloth datasets. From Table~\ref{tab:overall}, we have the following observations:

(1) BGAFlow consistently achieves the best results across all domains and evaluation metrics, with relative improvements ranging from 3.91\% to 16.62\% over the strongest competing baselines. These results demonstrate the effectiveness of the proposed framework in cross-domain sequential recommendation, particularly its ability to integrate transferable semantic information through personalized and state-adaptive representation transfer.

(2) Compared with single-domain sequential recommendation methods, including SASRec, FEARec, and UniSRec, BGAFlow achieves substantial improvements across all domains. This indicates that incorporating useful information from the paired domain can enhance within-domain recommendation. While single-domain methods rely solely on interaction histories from the current domain, BGAFlow exploits complementary cross-domain semantic information through a personalized transfer process. By preserving domain-specific collaborative representations and progressively refining transferable semantic features, BGAFlow effectively leverages cross-domain preferences to improve recommendation quality.

(3) BGAFlow also outperforms conventional cross-domain recommendation methods, including $\pi$-Net, C$^2$DSR, Tri-CDR, RecGURU, SyNCRec, and ABXI. Although these methods capture cross-domain correlations through representation sharing, interaction modeling, or alignment objectives, their transfer mechanisms differ in how explicitly they model user-specific transfer extent and intermediate adaptation states. In contrast, BGAFlow combines a personalized transfer horizon with a B\'ezier-guided state-adaptive residual flow, enabling semantic representations to be progressively refined according to evolving preference feedback. The superior performance suggests the potential benefits of jointly modeling transfer strength and state-dependent semantic adaptation.

(4) Compared with diffusion-based recommendation methods, including DiffuRec, DiffCDR, CD-CDR, and D2TCDR, BGAFlow also achieves consistently better performance. Unlike diffusion-based representation generation~\cite{liang2026semdiff}, BGAFlow establishes an explicit B\'ezier reference trajectory and learns task-aware residual corrections through a conditional velocity field. By incorporating intermediate-state preference feedback and smoothly integrating residual updates, the proposed framework provides a structured and adaptive mechanism for cross-domain semantic transfer. These results demonstrate the effectiveness of this approach under the evaluated recommendation settings.

(5) BGAFlow further surpasses recent LLM-based baselines, including LLM-ESR, LeCDSR, and LLM4CDSR. Although these methods benefit from LLM-enhanced semantic representations, their primary emphasis is on enriching semantic information or facilitating cross-domain knowledge transfer through semantic modeling. BGAFlow additionally leverages LLM-inferred user profiles to estimate user-specific preference overlap and guide a conditional B\'ezier reference, while state-adaptive residual flow dynamically incorporates prototype-based preference feedback during transfer. Together with task-guided velocity supervision, this design enables recommendation-aware semantic refinement beyond static semantic representation enhancement. The results highlight the effectiveness of combining LLM-derived preference knowledge with personalized and adaptive cross-domain transfer.

\begin{table}
\centering
\caption{Ablation studies of BGAFlow on Movie-Book and Sport-Cloth datasets. The best and second-best results in each column are boldfaced and underlined, respectively.}
\label{tab:ablation}
\vspace{-2mm}
\setlength{\tabcolsep}{2.2pt}
\renewcommand{\arraystretch}{1.05}
\footnotesize
\resizebox{0.98\columnwidth}{!}{
\begin{tabular}{lcccccccc}
\toprule
\multirow{2}{*}{Variants}
& \multicolumn{2}{c}{\textbf{Movie}}
& \multicolumn{2}{c}{\textbf{Book}}
& \multicolumn{2}{c}{\textbf{Sport}}
& \multicolumn{2}{c}{\textbf{Cloth}} \\
\cmidrule(lr){2-3}
\cmidrule(lr){4-5}
\cmidrule(lr){6-7}
\cmidrule(lr){8-9}
& HR@10 & NDCG@10
& HR@10 & NDCG@10
& HR@10 & NDCG@10
& HR@10 & NDCG@10 \\
\midrule
w/o Pretraining
& 0.4643 & 0.3559
& 0.3817 & 0.2714
& {\ul 0.2689} & {\ul 0.1775}
& {\ul 0.2403} & {\ul 0.1654} \\

Linear Flow
& 0.5171 & 0.4032
& 0.3676 & 0.2729
& 0.2280 & 0.1559
& 0.2150 & 0.1503 \\

w/o Smoothness
& {\ul 0.5490} & {\ul 0.4258}
& 0.3944 & \textbf{0.2951}
& 0.2685 & {\ul 0.1775}
& 0.2378 & 0.1631 \\

w/o Task-guided FM
& 0.5449 & 0.4206
& {\ul 0.3992} & 0.2945
& 0.2620 & 0.1723
& 0.2329 & 0.1622 \\

w/o dynamic feedback
& 0.5487 & 0.4242
& 0.3930 & 0.2942
& 0.2660 & 0.1766
& 0.2378 & 0.1641 \\

\rowcolor{gray!12} \textbf{BGAFlow}
& \textbf{0.5532} & \textbf{0.4333}
& \textbf{0.4002} & {\ul 0.2948}
& \textbf {0.2701} & \textbf{0.1799}
& \textbf{0.2460} & \textbf{0.1708} \\
\bottomrule
\end{tabular}
}
\vspace{-2mm}
\end{table}
\subsection{Further Analysis}

\subsubsection{\textbf{Ablation Study}}

To investigate the contributions of the key components in BGAFlow, we compare the full model with the following ablation variants:
\begin{itemize}[leftmargin=*]
    \item \textbf{w/o Pretraining} removes unified dual-domain pretraining and trains the proposed state-adaptive flow framework from scratch.
    \item \textbf{Linear Flow} replaces the conditional B\'ezier reference with a linear interpolation path while retaining the state-adaptive residual flow.
    \item \textbf{w/o Smoothness} removes the full-trajectory smoothness regularization while retaining task-guided flow matching and smoothstep-based integration.
    \item \textbf{w/o Task-guided FM} removes the task-guided flow matching objective and optimizes the residual flow using the recommendation and smoothness losses.
    \item \textbf{w/o dynamic feedback} replaces the dynamically updated prototype preference feedback with static feedback computed from the target-domain semantic representation.
\end{itemize}

Table~\ref{tab:ablation} reports HR@10 and NDCG@10 across four domains. BGAFlow achieves the best results on seven of the eight metrics and the second-best result on Book NDCG@10. Removing unified dual-domain pretraining reduces performance on all eight metrics, with the largest relative decreases on Movie, where HR@10 falls from 0.5532 to 0.4643 and NDCG@10 from 0.4333 to 0.3559. The smaller decreases on Sport indicate that the contribution of pretraining varies across domains. Replacing the conditional B\'ezier reference with linear interpolation also lowers all eight metrics, with particularly large reductions on Sport and Cloth. These results support the value of both pretrained semantic representations and a condition-guided geometric reference for cross-domain transfer.

Removing dynamic preference feedback degrades performance on every metric, suggesting that updating prototype feedback during semantic evolution helps the residual flow respond to intermediate states. Removing task-guided flow matching likewise lowers all eight metrics. Its effect is more pronounced on Movie and Sport-Cloth than on Book, where HR@10 and NDCG@10 decrease only slightly, from 0.4002 to 0.3992 and from 0.2948 to 0.2945, respectively. This pattern supports the contribution of recommendation-aware velocity supervision while showing that its benefit is domain-dependent.

Removing trajectory smoothness regularization lowers seven metrics, but slightly improves Book NDCG@10 from 0.2948 to 0.2951. Thus, smoothness regularization is generally beneficial, although its effect is not uniform across all domains and metrics. Taken together, the results support the contributions of BGAFlow's components to its overall performance without implying that every component improves every individual metric.

\subsubsection{\textbf{Pretraining Loss Ablation}}
\label{sec:loss_ablation}

\begin{figure}
    \centering
    \includegraphics[width=0.92\linewidth,height=0.31\textheight,keepaspectratio]{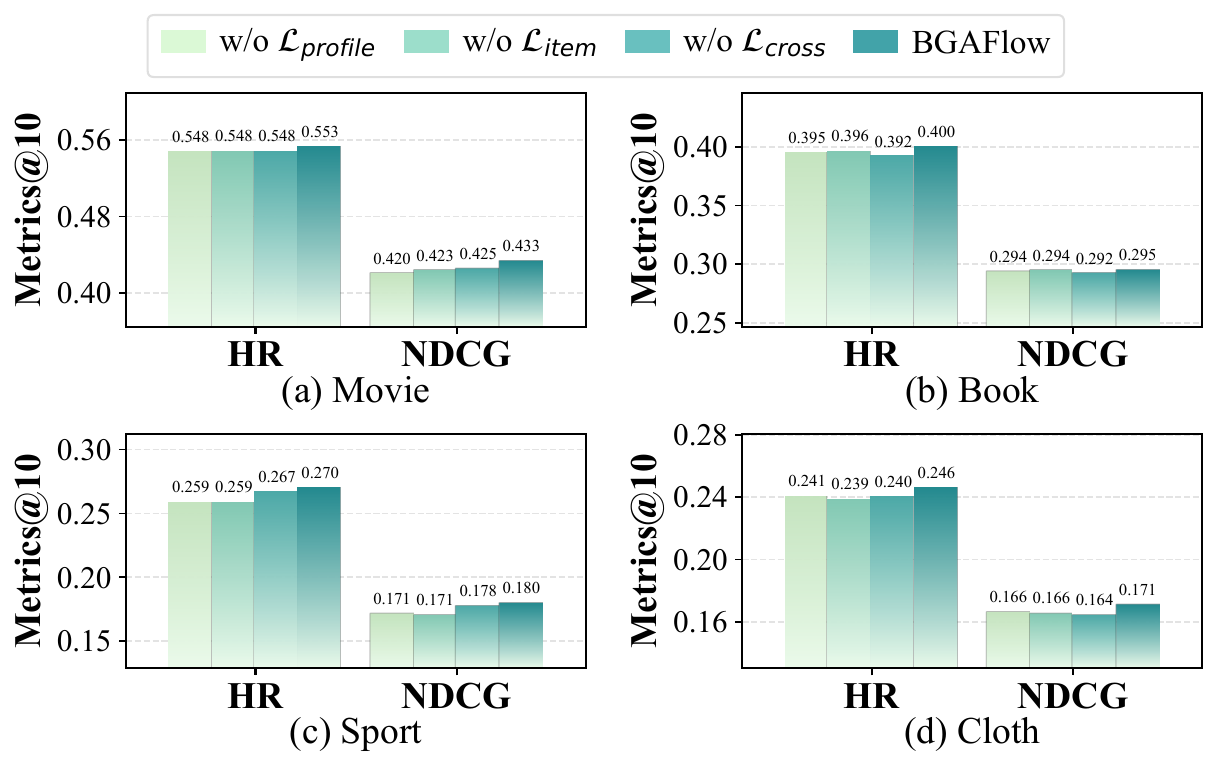}
    \caption{Study on auxiliary loss ablation in unified dual-domain pretraining.}
    \Description{Four panels show Movie, Book, Sport, and Cloth. Each panel contains grouped bars for HR at 10 and NDCG at 10, comparing the complete model with removal of profile alignment, item alignment, or cross-domain regularization. Numeric scores appear above the bars.}
    \label{fig:preloss-ablation}
\end{figure}

Fig.~\ref{fig:preloss-ablation} reports the ablation results of different pretraining objectives. Removing $\mathcal{L}_{\mathrm{profile}}$ leads to performance degradation in most domains, suggesting that LLM-inferred user profiles provide useful high-level preference supervision for semantic representation learning. Removing $\mathcal{L}_{\mathrm{item}}$ also worsens performance in most cases, indicating that aligning user semantic representations with ground-truth item semantics improves their relevance to next-item prediction. Furthermore, removing $\mathcal{L}_{\mathrm{cross}}$ causes noticeable performance drops, particularly on Book and Cloth, highlighting the importance of cross-domain distribution regularization in establishing a comparable semantic space for subsequent state-adaptive transfer.

\subsubsection{\textbf{Hyperparameter Sensitivity}}\label{sec:param_sensitivity}
\begin{figure}
    \centering

    \begin{minipage}[t]{0.25\linewidth}
        \centering
        \includegraphics[width=\linewidth]{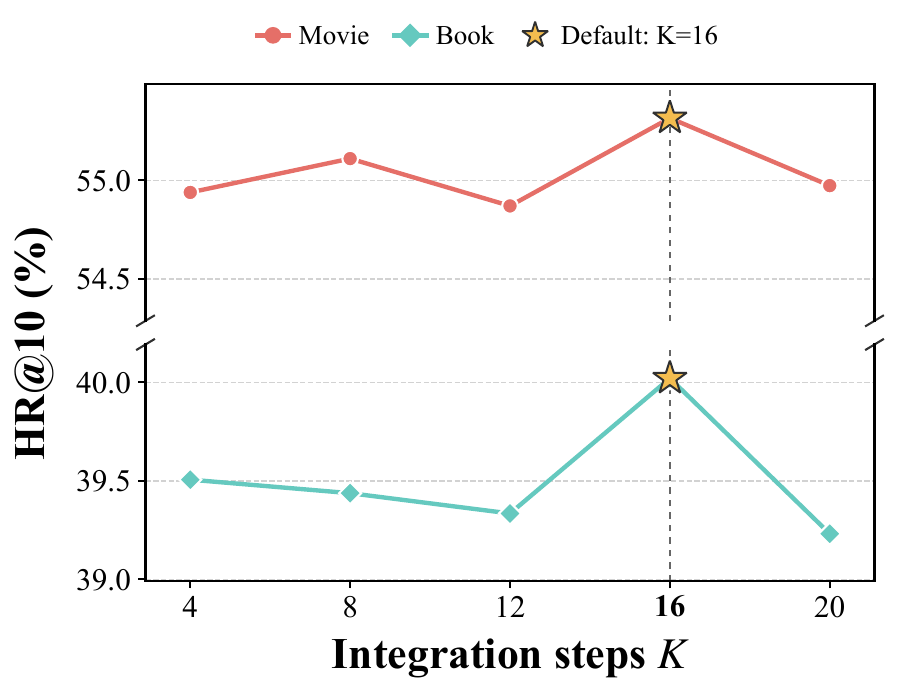}
        \vspace{-1.5mm}
        
        {\scriptsize (a) Movie-Book, $K$}
    \end{minipage}\hfill
    \begin{minipage}[t]{0.25\linewidth}
        \centering
        \includegraphics[width=\linewidth]{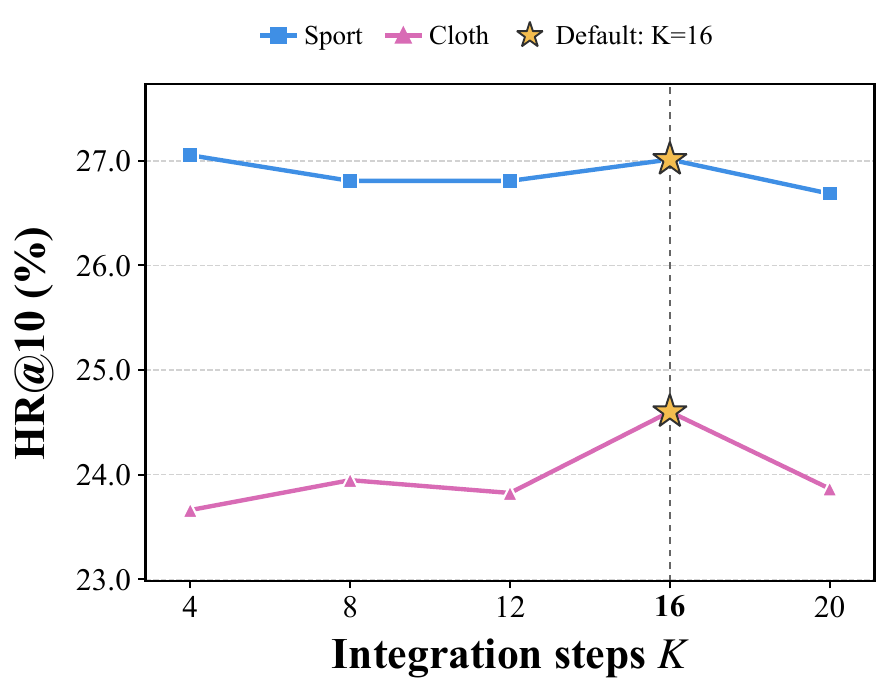}
        \vspace{-1.5mm}
        
        {\scriptsize (b) Sport-Cloth, $K$}
    \end{minipage}\hfill
    \begin{minipage}[t]{0.25\linewidth}
        \centering
        \includegraphics[width=\linewidth]{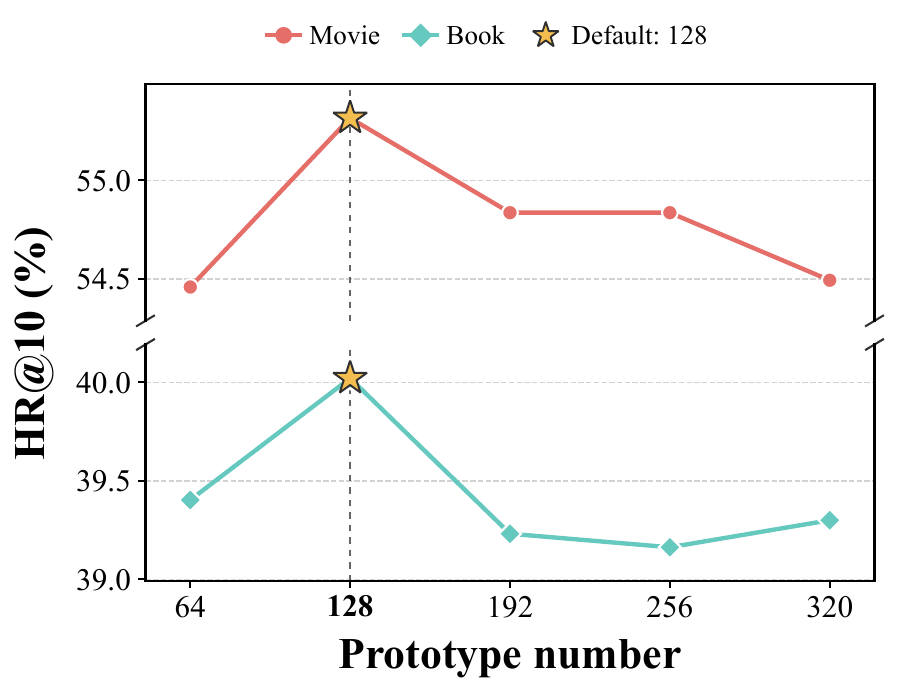}
        \vspace{-1.5mm}
        
        {\scriptsize (c) Movie-Book, $n_p$}
    \end{minipage}\hfill
    \begin{minipage}[t]{0.25\linewidth}
        \centering
        \includegraphics[width=\linewidth]{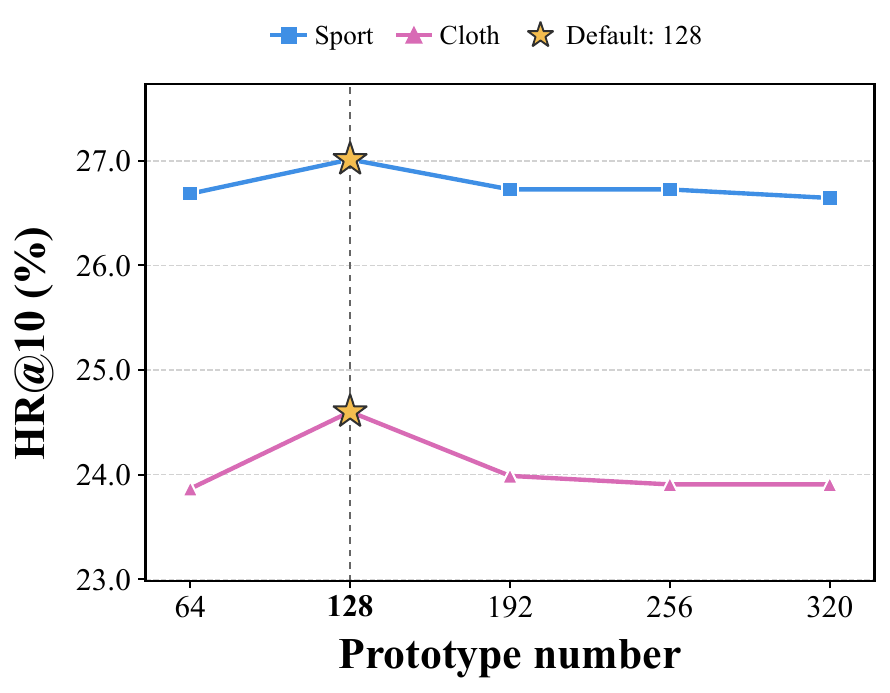}
        \vspace{-1.5mm}
        
        {\scriptsize (d) Sport-Cloth, $n_p$}
    \end{minipage}

    \caption{Impact of residual integration steps $K$ and prototype number $n_p$ on Movie-Book and Sport-Cloth datasets.}
    \Description{Four line charts use broken vertical axes to compare HR at 10 across integration-step counts and prototype counts. Each panel shows both domains in a dataset pair. Star markers identify the default settings of 16 integration steps and 128 prototypes.}
    \label{fig:param-study}
\end{figure}

Fig.~\ref{fig:param-study} investigates the effects of the number of residual flow integration steps $K$ and the number of latent preference prototypes $n_p$. For $K$, BGAFlow generally achieves its best performance at $K=16$, except for Sport, where a slightly higher HR@10 is observed at $K=4$. In the proposed framework, $K$ determines the granularity of state-adaptive residual integration, allowing the velocity network to progressively refine intermediate representations using dynamically updated preference feedback. However, increasing $K$ does not consistently improve performance and introduces additional computational overhead, suggesting that a moderate number of integration steps is sufficient for effective semantic adaptation. For $n_p$, the prototypes support both personalized horizon estimation and dynamic preference feedback. As shown in Fig.~\ref{fig:param-study}, $n_p=128$ achieves the best performance across all four domains. Too few prototypes may limit the representation of fine-grained user interests, whereas increasing their number does not necessarily provide more useful preference information. Based on the overall validation performance, we adopt $K=16$ and $n_p=128$ as the default settings.

\subsubsection{\textbf{Study on Horizon Estimation}}\label{sec:ex_horizon}

\begin{figure}
\setlength{\abovecaptionskip}{0.0cm}
\setlength{\belowcaptionskip}{0.0cm} 
    \centering
    \includegraphics[width=0.8\textwidth]{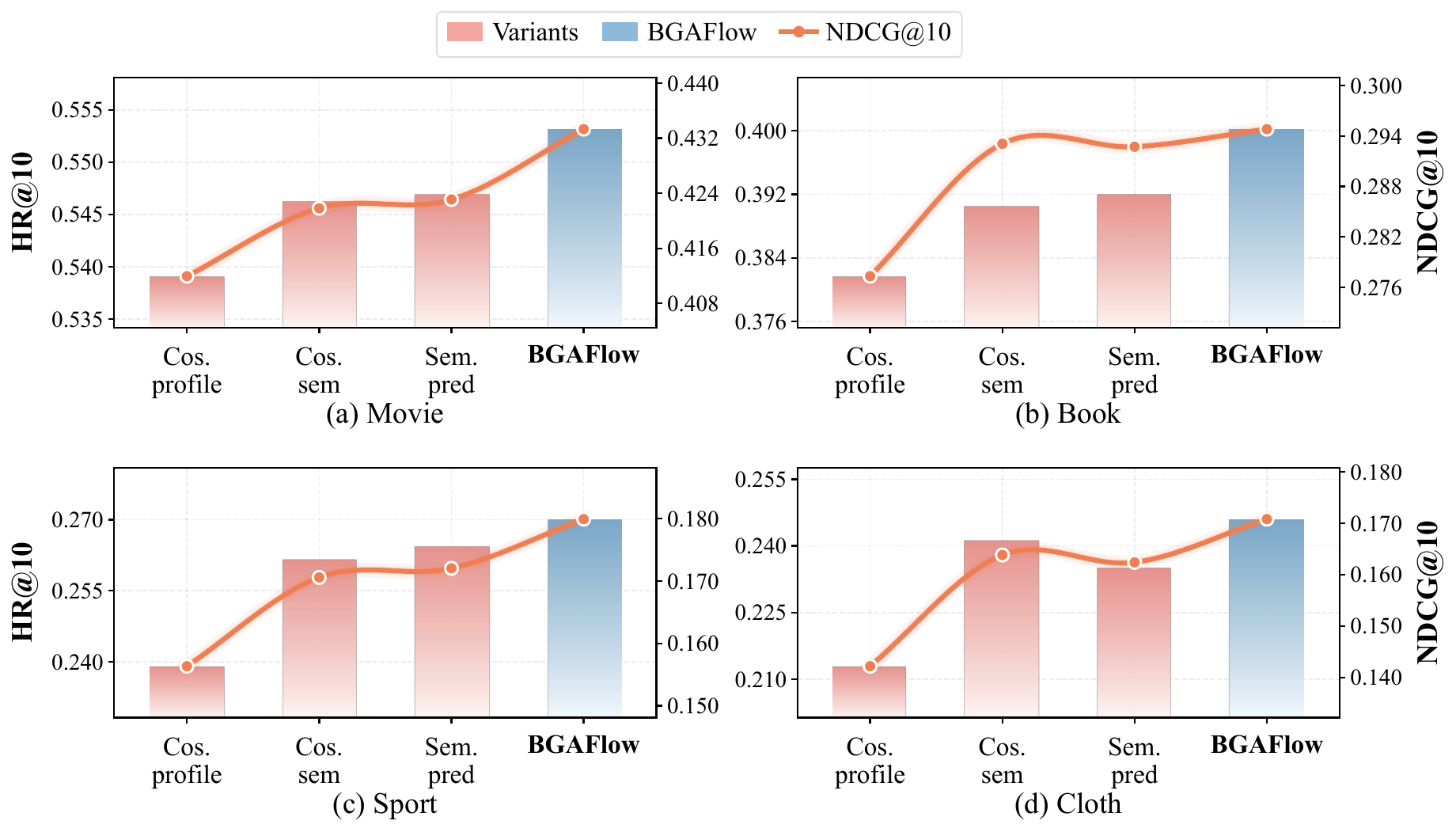}
    \caption{Comparison of horizon prediction methods on different datasets.}
    \Description{Four panels compare cosine profile similarity, cosine semantic similarity, a semantic horizon predictor, and BGAFlow on Movie, Book, Sport, and Cloth. Bars show HR at 10 and lines show NDCG at 10, with separate vertical axes for the two metrics.}
\label{fig:horizon_variant}
\end{figure}
To evaluate the effectiveness of personalized horizon estimation in the proposed B\'ezier-guided state-adaptive flow, we compare BGAFlow with three variants. \textbf{Cos. profile} directly uses the cosine similarity between two domain-level user profile embeddings as the horizon signal. \textbf{Cos. sem} uses the cosine similarity between the learned semantic representations of the two domains. \textbf{Sem. pred} retains the horizon predictor structure of BGAFlow but replaces profile-based inputs with learned semantic representations. In contrast, BGAFlow combines prototype-level preference overlap and condition-related geometric similarities to estimate the personalized horizon, which controls both the effective B\'ezier reference extent and the residual correction scale.

As shown in Fig.~\ref{fig:horizon_variant}, BGAFlow consistently achieves the best HR@10 and NDCG@10 across all four domains, demonstrating the effectiveness of the proposed horizon estimation strategy. The inferior performance of \textbf{Cos. profile} suggests that direct profile similarity may be insufficient to capture fine-grained transferable interests. \textbf{Cos. sem} also performs worse, potentially because the learned semantic representations are jointly optimized and their similarity may not reliably reflect user-specific transferability. Although \textbf{Sem. pred} improves over simple cosine-based variants in several cases, it remains inferior to BGAFlow. These results suggest that combining prototype-level preference overlap with condition-related geometric information provides more informative signals for controlling personalized B\'ezier-guided semantic adaptation.

\begin{figure}
\setlength{\abovecaptionskip}{0.0cm}
\setlength{\belowcaptionskip}{0.0cm} 
    \centering
    \includegraphics[width=0.85\textwidth]{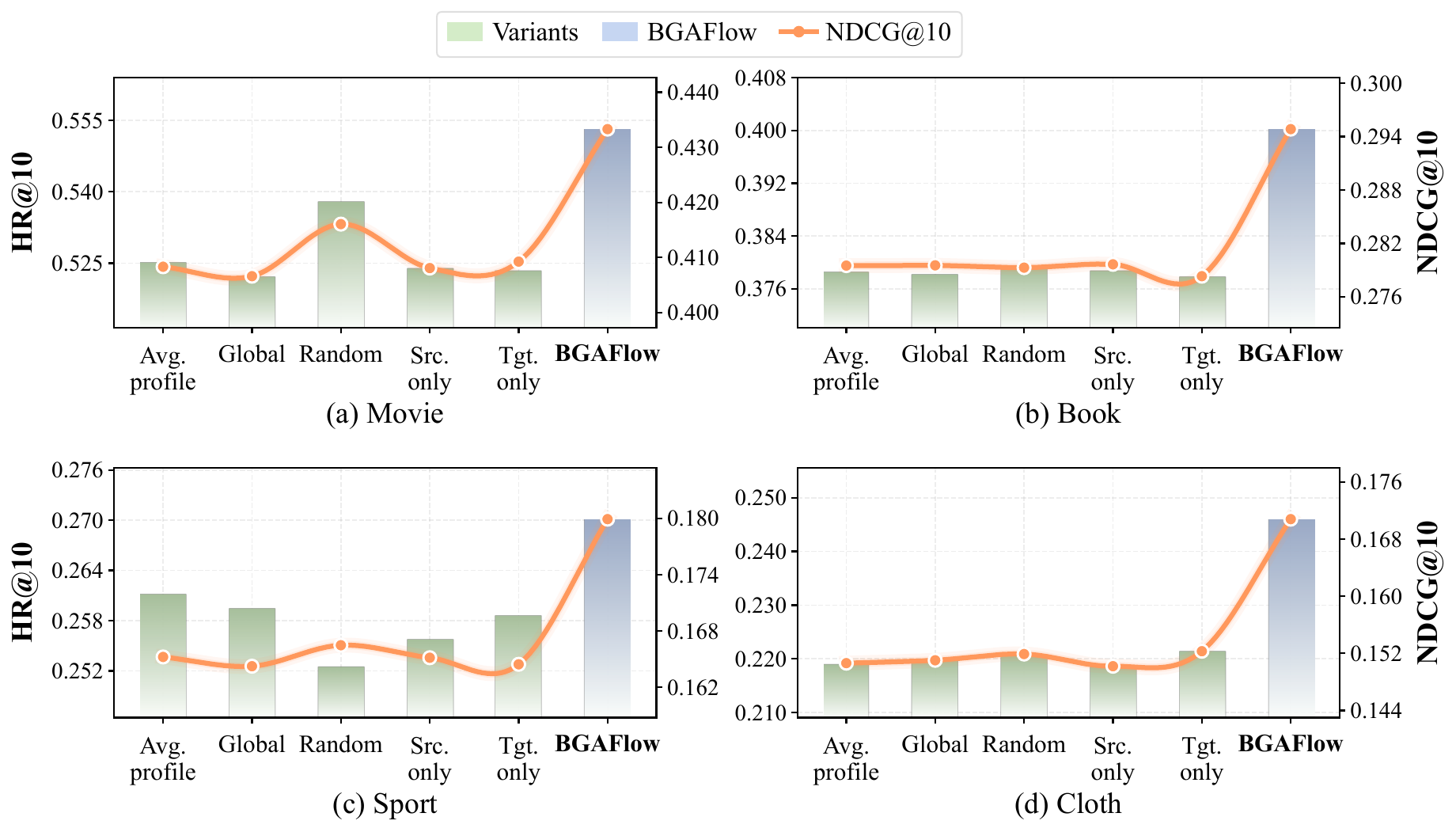}
    \caption{Comparison of conditional B\'ezier control-point configurations across four recommendation domains.}
    \Description{Four panels compare averaged profiles, a global control point, a random control point, source-only profiles, target-only profiles, and BGAFlow on Movie, Book, Sport, and Cloth. Bars show HR at 10 and lines show NDCG at 10.}
\label{fig:waypoint_variant}
\end{figure}
\subsubsection{\textbf{Study on B\'ezier Trajectory Waypoint}}\label{sec:ex_waypoint}
To investigate the effect of the conditional B\'ezier control point, we compare BGAFlow with five variants. \textbf{Avg. profile} uses the weighted average of two domain-level user profile embeddings as the control point. \textbf{Global} adopts a single learnable control point shared by all users, while \textbf{Random} uses a randomly initialized control point without user-specific semantic guidance. \textbf{Src. only} and \textbf{Tgt. only} construct the control point using only the source-domain and target-domain profile embeddings, respectively. In contrast, BGAFlow learns a user-conditioned control point by jointly incorporating preference information from both domains.

As shown in Fig.~\ref{fig:waypoint_variant}, BGAFlow consistently achieves the best HR@10 and NDCG@10 across all four domains, demonstrating the effectiveness of its condition-guided control point. The inferior performance of \textbf{Global} and \textbf{Random} suggests that user-specific semantic guidance is important for constructing an appropriate geometric reference for cross-domain transfer. Compared with \textbf{Avg. profile}, BGAFlow benefits from learning a more flexible fusion of cross-domain preferences, which jointly conditions the B\'ezier reference and residual velocity field. Moreover, the weaker results of \textbf{Src. only} and \textbf{Tgt. only} indicate that relying on one-sided preference information may limit the ability to capture transferable semantic relationships. Overall, these results highlight the importance of integrating complementary cross-domain preferences to guide personalized and state-adaptive semantic refinement.

\begin{figure}
    \centering
    \subfigure[]{%
        \includegraphics[width=0.4\linewidth]{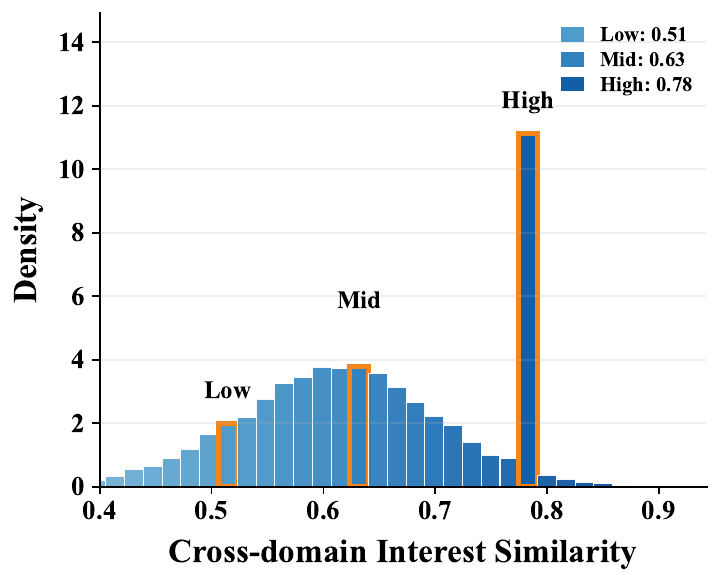}%
        \label{fig:interest_overlap_mb}%
    }\hfill
    \subfigure[]{%
        \includegraphics[width=0.4\linewidth]{Fig/SC_interest_overlap.pdf}%
        \label{fig:interest_overlap_sc}%
    }
    \caption{Distributions of cross-domain interest similarity on (a) Movie-Book and (b) Sport-Cloth.}
    \Description{Two histograms show the density of user interest similarity. Movie-Book has a broad distribution and a prominent peak near 0.78. Sport-Cloth is concentrated in the middle-to-high similarity range, with a prominent peak near 0.78.}
    \label{fig:interest_overlap}
\end{figure}

\begin{figure*}
    \centering
    \includegraphics[width=\columnwidth]{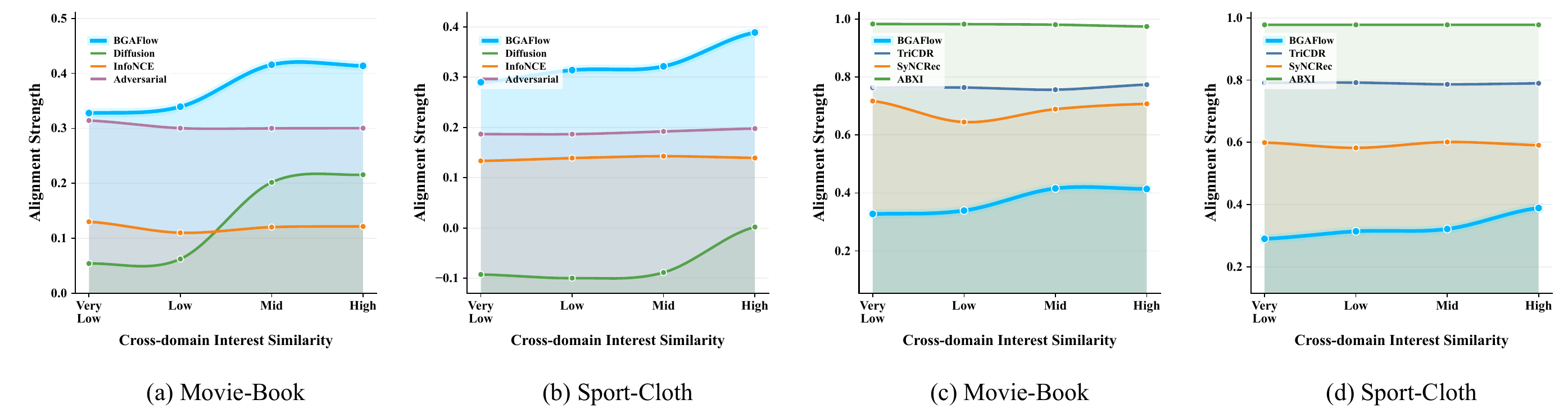}
   \caption{Alignment strength across cross-domain interest similarity groups: (a)--(b) compare different alignment variants, and (c)--(d) compare different baseline methods. Panels (a) and (c) show Movie-Book; panels (b) and (d) show Sport-Cloth.}
    \Description{Four line charts compare alignment strength across Very Low, Low, Mid, and High interest similarity groups. The first two panels compare alignment variants on Movie-Book and Sport-Cloth. The last two compare BGAFlow with Tri-CDR, SyNCRec, and ABXI on the same dataset pairs.}
\label{fig:align_strength}
\end{figure*}

\begin{figure*}
    \centering
    \includegraphics[width=\columnwidth]{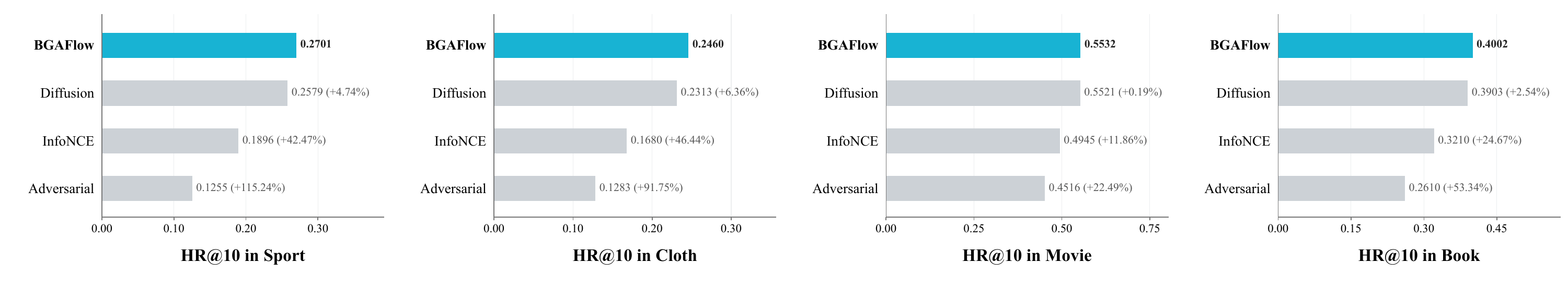}
    \caption{HR@10 comparison of different cross-domain alignment mechanisms across four recommendation domains.}
    \Description{Four horizontal bar charts compare BGAFlow, Diffusion, InfoNCE, and adversarial alignment using HR at 10 on Sport, Cloth, Movie, and Book. BGAFlow has the highest reported score in each domain.}
\label{fig:align_performance}
\end{figure*}

\subsubsection{\textbf{Analysis of Dynamic Alignment Behavior}}
\label{sec:ex_align}

We investigate the dynamic alignment behavior of BGAFlow on Movie-Book and Sport-Cloth through cross-domain interest similarity distributions, alignment strength, and recommendation performance. Users are categorized into Very Low, Low, Mid, and High groups according to the interest similarity estimated from LLM-inferred profiles. This grouping provides a diagnostic measure of cross-domain preference consistency, while BGAFlow estimates personalized transfer horizons using prototype-level preference overlap and condition-related geometric information.

Fig.~\ref{fig:interest_overlap} illustrates the distributions of cross-domain interest similarity. Movie-Book exhibits a relatively broad distribution with a noticeable concentration of high-similarity users, while Sport-Cloth shows a more concentrated distribution in the middle-to-high similarity range. These heterogeneous preference relationships highlight the importance of personalized transfer control, as users with different degrees of cross-domain interest consistency may require different levels of semantic adaptation.

Fig.~\ref{fig:align_strength}(a)--(b) compares BGAFlow with different alignment mechanisms using the same backbone and semantic features, where Diffusion also adopts BGAFlow's horizon predictor. InfoNCE and Adversarial alignment exhibit relatively stable alignment strengths across different similarity groups, suggesting limited sensitivity to user-specific preference consistency. In contrast, BGAFlow generally produces stronger alignment for users with higher interest similarity, with noticeable increases from Low to Mid on Movie-Book and from Mid to High on Sport-Cloth. Diffusion also exhibits similarity-dependent variations, particularly on Movie-Book, indicating that alignment behavior is influenced by the transfer mechanism beyond horizon estimation alone. In BGAFlow, the personalized horizon controls the B\'ezier reference extent and residual correction scale, while the state-adaptive flow progressively refines the transfer trajectory using intermediate preference feedback. The observed alignment patterns are consistent with this personalized and adaptive transfer design.

Fig.~\ref{fig:align_strength}(c)--(d) further compares BGAFlow with Tri-CDR, SyNCRec, and ABXI. These baselines generally maintain relatively high alignment strengths across user groups. In particular, Tri-CDR and ABXI exhibit nearly unchanged alignment levels, while SyNCRec shows variations that do not consistently correspond to increasing interest similarity. By comparison, BGAFlow exhibits clearer differences between low- and high-similarity groups, suggesting greater sensitivity to user-specific preference relationships. This observation is important because stronger alignment does not necessarily imply more effective knowledge transfer when users exhibit weak cross-domain preference correspondence.

Fig.~\ref{fig:align_performance} reports the HR@10 results of different alignment variants across four domains. BGAFlow achieves the best performance, reaching 0.2701 on Sport, 0.2460 on Cloth, 0.5532 on Movie, and 0.4002 on Book. Compared with Diffusion, the strongest competing variant, BGAFlow achieves relative improvements of 4.74\%, 6.36\%, 0.19\%, and 2.54\%, respectively, with nearly comparable performance on Movie. More substantial improvements are observed over InfoNCE and Adversarial alignment. Together with the alignment behavior analysis, these results support the effectiveness of combining personalized transfer horizons with B\'ezier-guided state-adaptive residual refinement, while indicating that the magnitude of recommendation gains varies across domains.

\begin{figure*}
    \centering
    \includegraphics[width=\columnwidth]{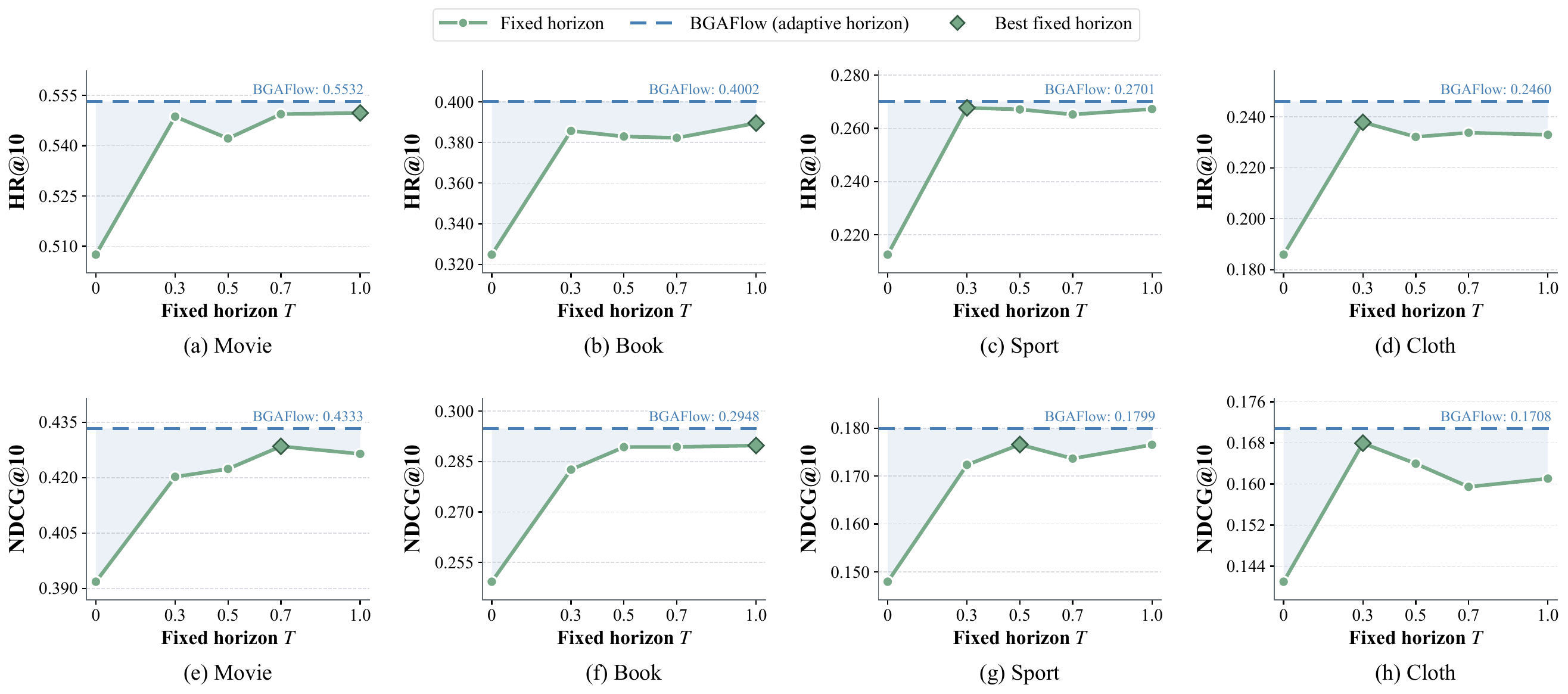}
 \caption{Performance comparison between BGAFlow with personalized transfer horizons and fixed-horizon variants across four recommendation domains. The dashed lines indicate the performance of BGAFlow, and the diamond markers denote the best results among the evaluated fixed-horizon settings.}
\Description{Eight panels show four domains in columns and HR at 10 and NDCG at 10 in rows. Each panel compares performance at five fixed transfer horizons with the horizontal dashed BGAFlow score. Diamond markers identify the best evaluated fixed horizon.}
\label{fig:fixed_horizon}
\end{figure*}
\subsubsection{\textbf{Adaptive versus Fixed Transfer Horizons}}

To further investigate the necessity of personalized transfer horizons, we compare BGAFlow with fixed-horizon variants by replacing the user-specific horizon $T_i$ with a globally shared value $T \in \{0, 0.3, 0.5, 0.7, 1.0\}$ while retaining the remaining model components. In the proposed framework, the horizon controls both the effective extent of the B\'ezier reference and the scale of residual flow corrections. This comparison examines whether user-specific horizon estimation provides additional benefits beyond selecting a fixed transfer extent.

As shown in Fig.~\ref{fig:fixed_horizon}, BGAFlow consistently outperforms all evaluated fixed-horizon settings across four domains and both evaluation metrics. Although increasing $T$ from zero generally improves recommendation performance, further increases do not consistently yield additional gains. Moreover, the best-performing fixed horizon varies across domains and evaluation metrics, indicating that a single transfer extent cannot consistently accommodate heterogeneous cross-domain preference relationships. Notably, even the best fixed-horizon results remain below those of BGAFlow, although the performance gaps are relatively small in certain settings, such as Sport. These results provide further evidence that personalized horizon estimation enables more effective control over B\'ezier-guided semantic adaptation than relying on a globally fixed transfer extent.

\begin{figure*}
    \centering
    \includegraphics[width=\columnwidth]{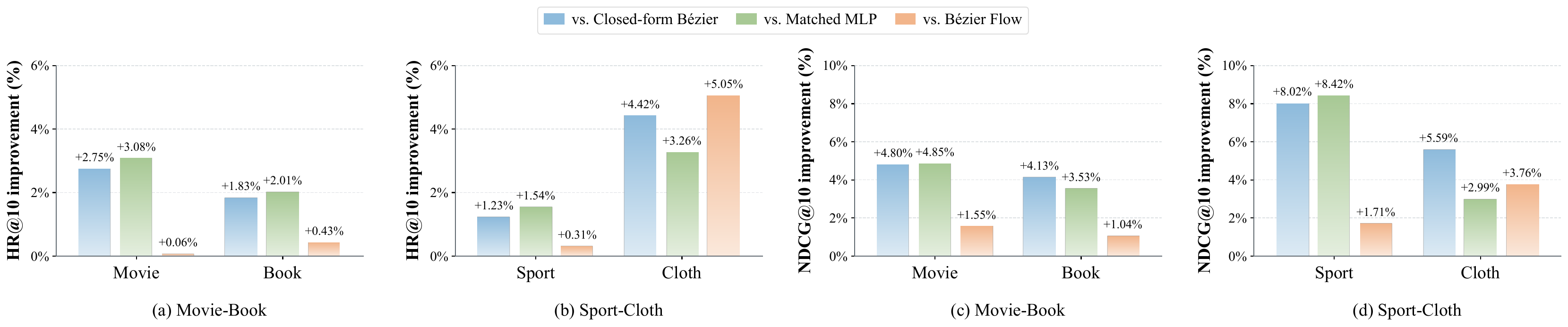}
\caption{Relative improvements of BGAFlow over three alternative transfer mechanisms across four recommendation domains. Panels (a) and (b) report HR@10 improvements on Movie-Book and Sport-Cloth, respectively, while panels (c) and (d) report the corresponding NDCG@10 improvements.}
\Description{Four grouped bar charts show the percentage improvement of BGAFlow over closed-form Bezier transfer, a matched MLP, and Bezier Flow. The first two panels show HR at 10 and the last two show NDCG at 10, each split into Movie-Book and Sport-Cloth.}
\label{fig:flow_necessity}
\end{figure*}

\subsubsection{\textbf{Necessity of State-Adaptive Residual Flow}}
\label{sec:flow_necessity}

To investigate the necessity of the proposed state-adaptive residual flow, we compare BGAFlow with three alternative transfer mechanisms. \textbf{Closed-form B\'ezier} directly computes the adapted representation from the conditional B\'ezier reference without learning residual corrections. \textbf{Matched MLP} introduces a parameter-matched MLP to predict a one-step correction to the B\'ezier reference, replacing iterative flow-based adaptation. \textbf{B\'ezier Flow} follows the original flow matching formulation, where the velocity network learns to approximate the analytical B\'ezier tangent for representation transfer. In contrast, BGAFlow employs task-guided residual velocity learning and dynamically updated preference feedback to progressively refine the B\'ezier reference.

Fig.~\ref{fig:flow_necessity} reports the relative improvements of BGAFlow over these variants across four domains. BGAFlow consistently outperforms all three alternatives on both HR@10 and NDCG@10. Compared with \textbf{Closed-form B\'ezier}, BGAFlow achieves improvements of up to 4.42\% in HR@10 and 8.02\% in NDCG@10, indicating the benefits of learning task-adaptive corrections beyond a predetermined geometric reference. BGAFlow also surpasses \textbf{Matched MLP} across all domains, suggesting that progressive residual refinement provides advantages over one-step correction. Furthermore, compared with \textbf{B\'ezier Flow}, BGAFlow achieves consistent NDCG@10 improvements ranging from 1.04\% to 3.76\%, although the HR@10 gains are relatively modest in some domains. These results highlight the effectiveness of combining structured B\'ezier guidance with task-aware, state-adaptive residual evolution, rather than relying solely on analytical interpolation or direct velocity matching.

\begin{figure*}
    \centering
    \includegraphics[width=\columnwidth]{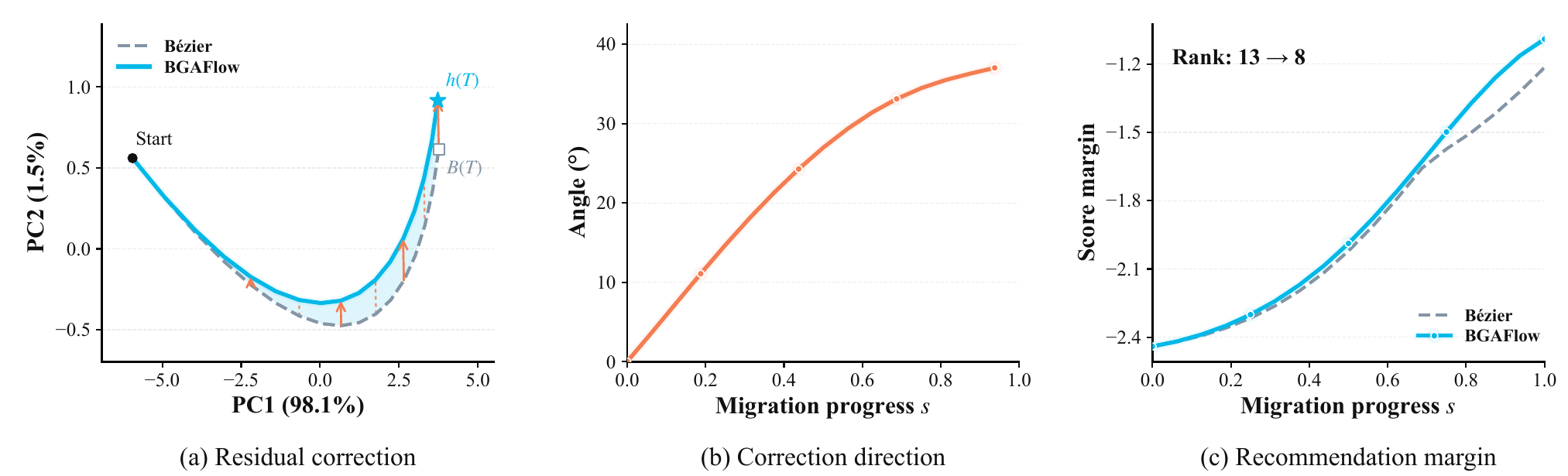}
\caption{Case study of B\'ezier-guided state-adaptive residual refinement on Sport. (a) PCA visualization of the analytical B\'ezier reference and the learned BGAFlow trajectory, with arrows indicating residual corrections. (b) Angular deviation of the residual velocity direction from its initial direction over migration progress. (c) Evolution of the target-item score margin against the highest-scoring negative item, where BGAFlow improves the target item's rank from 13 to 8. The direction angles and score margins are computed in the original 128-dimensional representation space.}
\Description{Three panels show a Sport-domain case study: reference and adapted trajectories in two principal components, residual direction changes reaching about 37 degrees, and target-item score margins during transfer. The final target rank changes from 13 for the reference to 8 for BGAFlow.}
\label{fig:case_study}

\end{figure*}

\subsubsection{\textbf{Case Study of State-Adaptive Residual Refinement}}
\label{sec:case_study}

To provide an intuitive understanding of how the proposed state-adaptive residual flow refines cross-domain semantic transfer, we visualize the migration process of a selected user in the Sport domain. As shown in Fig.~\ref{fig:case_study}(a), the learned BGAFlow trajectory generally follows the B\'ezier geometric reference while introducing progressive residual corrections at intermediate states. These corrections gradually shift the actual trajectory away from the analytical reference, resulting in a different terminal representation. Although the trajectories are visualized through two-dimensional PCA, the residual corrections are learned and applied in the original 128-dimensional semantic space.

To examine how the learned correction direction evolves, Fig.~\ref{fig:case_study}(b) reports its angular deviation from the initial residual velocity. Let $\bm{g}_i^k$ denote the norm-clipped output of the velocity network at step $k$, before applying the smoothstep coefficient and residual scaling. For nonzero initial and current velocity directions, the angular deviation is defined as
\begin{equation}
\label{eq:case_angle}
\mathrm{Angle}_i^k
=
\frac{180}{\pi}
\arccos\left(
\mathrm{clip}_{[-1,1]}
\left(
\frac{
\langle \bm{g}_i^k,\bm{g}_i^0\rangle
}{
\|\bm{g}_i^k\|_2\|\bm{g}_i^0\|_2
}
\right)
\right).
\end{equation}
As migration progresses, the angular deviation gradually increases to approximately $37^\circ$, indicating that the learned residual direction changes relative to its initial orientation. This observation provides an example of state-dependent velocity adjustments during the transfer process, rather than a constant correction direction throughout the trajectory.

To further examine the recommendation effect of these adjustments, Fig.~\ref{fig:case_study}(c) compares the target-item score margins along the B\'ezier reference and BGAFlow trajectories. For a fixed candidate set containing one ground-truth item $v_i^+$ and 999 negative items $\mathcal{N}_i$, the margin at step $k$ is defined as
\begin{equation}
\label{eq:case_margin}
m_i^k
=
S_i^k(v_i^+)
-
\max_{v\in\mathcal{N}_i}S_i^k(v),
\end{equation}
where $S_i^k(\cdot)$ denotes the complete recommendation score combining collaborative and semantic predictions. The two trajectories share identical collaborative representations, item embeddings, gating weights, and candidate items, differing only in their intermediate semantic representations. As shown in Fig.~\ref{fig:case_study}(c), BGAFlow achieves a higher (less negative) score margin than the B\'ezier reference in the later migration stages, indicating an improved relative score for the target item against its strongest competing negative item. In this case, the target item's final rank also improves from 13 to 8, entering the Top-10 recommendation list. Together, these observations illustrate how state-adaptive residual corrections can refine the geometric reference and yield more favorable recommendation outcomes.

\section{Conclusion}
In this work, we address the limitations of static alignment in cross-domain sequential recommendation by proposing BGAFlow, a personalized framework for continuous and adaptive semantic transfer. We first establish a shared semantic space through unified dual-domain pretraining, which preserves domain-specific collaborative information while learning comparable semantic representations. To enable flexible knowledge transfer, BGAFlow constructs a conditional B\'ezier curve as a structured geometric reference and introduces a state-adaptive residual flow that progressively refines the transfer trajectory according to evolving preference states. Furthermore, a personalized horizon is estimated from latent preference overlap to control the extent of semantic transfer, while task-guided flow matching and trajectory smoothness regularization encourage effective and gradual representation adaptation. Extensive experiments on real-world cross-domain datasets demonstrate that BGAFlow consistently outperforms state-of-the-art baselines, validating the effectiveness of personalized transfer control and adaptive residual refinement for cross-domain recommendation.

\begin{acks}
This work is supported by the National Natural Science Foundation of China (No. 62272001 and No. 62206002), Natural Science Foundation of Anhui Province (No. 2208085QF195).
\end{acks}
\bibliographystyle{ACM-Reference-Format}
\bibliography{references}

@article{chen2024survey,
  title={A survey on cross-domain sequential recommendation},
  author={Chen, Shu and Xu, Zitao and Pan, Weike and Yang, Qiang and Ming, Zhong},
  journal={arXiv preprint arXiv:2401.04971},
  year={2024}
}

@article{zang2022survey,
  title={A survey on cross-domain recommendation: taxonomies, methods, and future directions},
  author={Zang, Tianzi and Zhu, Yanmin and Liu, Haobing and Zhang, Ruohan and Yu, Jiadi},
  journal={ACM Transactions on Information Systems},
  volume={41},
  number={2},
  pages={1--39},
  year={2022},
  publisher={ACM New York, NY}
}

@inproceedings{kang2018self,
  title={Self-attentive sequential recommendation},
  author={Kang, Wang-Cheng and McAuley, Julian},
  booktitle={2018 IEEE international conference on data mining (ICDM)},
  pages={197--206},
  year={2018},
  organization={IEEE}
}

@inproceedings{hu2018conet,
  title={{CoNet}: Collaborative cross networks for cross-domain recommendation},
  author={Hu, Guangneng and Zhang, Yu and Yang, Qiang},
  booktitle={Proceedings of the 27th ACM international conference on information and knowledge management},
  pages={667--676},
  year={2018}
}

@article{guo2021gcn,
  title={DA-GCN: A domain-aware attentive graph convolution network for shared-account cross-domain sequential recommendation},
  author={Guo, Lei and Tang, Li and Chen, Tong and Zhu, Lei and Nguyen, Quoc Viet Hung and Yin, Hongzhi},
  journal={arXiv preprint arXiv:2105.03300},
  year={2021}
}

@inproceedings{ma2019pi,
  title={$\pi$-net: A parallel information-sharing network for shared-account cross-domain sequential recommendations},
  author={Ma, Muyang and Ren, Pengjie and Lin, Yujie and Chen, Zhumin and Ma, Jun and Rijke, Maarten de},
  booktitle={Proceedings of the 42nd international ACM SIGIR conference on research and development in information retrieval},
  pages={685--694},
  year={2019}
}

@inproceedings{li2024cdrnp,
  title={CDRNP: Cross-domain recommendation to cold-start users via neural process},
  author={Li, Xiaodong and Sheng, Jiawei and Cao, Jiangxia and Zhang, Wenyuan and Li, Quangang and Liu, Tingwen},
  booktitle={Proceedings of the 17th ACM international conference on web search and data mining},
  pages={378--386},
  year={2024}
}

@article{ma2024triple,
  title={Triple sequence learning for cross-domain recommendation},
  author={Ma, Haokai and Xie, Ruobing and Meng, Lei and Chen, Xin and Zhang, Xu and Lin, Leyu and Zhou, Jie},
  journal={ACM Transactions on Information Systems},
  volume={42},
  number={4},
  pages={1--29},
  year={2024},
  publisher={ACM New York, NY}
}

@inproceedings{cao2022contrastive,
  title={Contrastive cross-domain sequential recommendation},
  author={Cao, Jiangxia and Cong, Xin and Sheng, Jiawei and Liu, Tingwen and Wang, Bin},
  booktitle={Proceedings of the 31st ACM international conference on information \& knowledge management},
  pages={138--147},
  year={2022}
}

@inproceedings{liu2025bridge,
  title={Bridge the domains: Large language models enhanced cross-domain sequential recommendation},
  author={Liu, Qidong and Zhao, Xiangyu and Wang, Yejing and Zhang, Zijian and Zhong, Howard and Chen, Chong and Li, Xiang and Huang, Wei and Tian, Feng},
  booktitle={Proceedings of the 48th International ACM SIGIR Conference on Research and Development in Information Retrieval},
  pages={1582--1592},
  year={2025}
}

@article{guo2023dan,
  title={DA-DAN: A dual adversarial domain adaption network for unsupervised non-overlapping cross-domain recommendation},
  author={Guo, Lei and Liu, Hao and Zhu, Lei and Guan, Weili and Cheng, Zhiyong},
  journal={ACM Transactions on Information Systems},
  volume={42},
  number={2},
  pages={1--27},
  year={2023},
  publisher={ACM New York, NY, USA}
}

@article{zhang2021deep,
  title={A deep dual adversarial network for cross-domain recommendation},
  author={Zhang, Qian and Liao, Wenhui and Zhang, Guangquan and Yuan, Bo and Lu, Jie},
  journal={IEEE Transactions on Knowledge and Data Engineering},
  volume={35},
  number={4},
  pages={3266--3278},
  year={2021},
  publisher={IEEE}
}

@inproceedings{li2022recguru,
  title={RecGURU: Adversarial learning of generalized user representations for cross-domain recommendation},
  author={Li, Chenglin and Zhao, Mingjun and Zhang, Huanming and Yu, Chenyun and Cheng, Lei and Shu, Guoqiang and Kong, Beibei and Niu, Di},
  booktitle={Proceedings of the fifteenth ACM international conference on web search and data mining},
  pages={571--581},
  year={2022}
}

@article{su2024cross,
  title={Cross-domain recommendation via dual adversarial adaptation},
  author={Su, Hongzu and Li, Jingjing and Du, Zhekai and Zhu, Lei and Lu, Ke and Shen, Heng Tao},
  journal={ACM Transactions on Information Systems},
  volume={42},
  number={3},
  pages={1--26},
  year={2024},
  publisher={ACM New York, NY}
}

@inproceedings{zhao2023cross,
  title={Cross-domain recommendation via user interest alignment},
  author={Zhao, Chuang and Zhao, Hongke and He, Ming and Zhang, Jian and Fan, Jianping},
  booktitle={Proceedings of the ACM web conference 2023},
  pages={887--896},
  year={2023}
}

@inproceedings{song2024mitigating,
  title={Mitigating negative transfer in cross-domain recommendation via knowledge transferability enhancement},
  author={Song, Zijian and Zhang, Wenhan and Deng, Lifang and Zhang, Jiandong and Wu, Zhihua and Bian, Kaigui and Cui, Bin},
  booktitle={Proceedings of the 30th ACM SIGKDD Conference on Knowledge Discovery and Data Mining},
  pages={2745--2754},
  year={2024}
}

@inproceedings{park2023cracking,
  title={Cracking the code of negative transfer: A cooperative game theoretic approach for cross-domain sequential recommendation},
  author={Park, Chung and Kim, Taesan and Choi, Taekyoon and Hong, Junui and Yu, Yelim and Cho, Mincheol and Lee, Kyunam and Ryu, Sungil and Yoon, Hyungjun and Choi, Minsung and others},
  booktitle={Proceedings of the 32nd ACM international conference on information and knowledge management},
  pages={2024--2033},
  year={2023}
}

@article{lipman2022flow,
  title={Flow matching for generative modeling},
  author={Lipman, Yaron and Chen, Ricky TQ and Ben-Hamu, Heli and Nickel, Maximilian and Le, Matt},
  journal={arXiv preprint arXiv:2210.02747},
  year={2022}
}

@article{liu2022flow,
  title={Flow straight and fast: Learning to generate and transfer data with rectified flow},
  author={Liu, Xingchao and Gong, Chengyue and Liu, Qiang},
  journal={arXiv preprint arXiv:2209.03003},
  year={2022}
}

@inproceedings{liu2022exploiting,
  title={Exploiting variational domain-invariant user embedding for partially overlapped cross domain recommendation},
  author={Liu, Weiming and Zheng, Xiaolin and Su, Jiajie and Hu, Mengling and Tan, Yanchao and Chen, Chaochao},
  booktitle={Proceedings of the 45th International ACM SIGIR conference on research and development in information retrieval},
  pages={312--321},
  year={2022}
}

@inproceedings{bian2025abxi,
  title={{ABXI}: invariant interest adaptation for task-guided cross-domain sequential recommendation},
  author={Bian, Qingtian and de Carvalho, Marcus and Li, Tieying and Xu, Jiaxing and Fang, Hui and Ke, Yiping},
  booktitle={Proceedings of the ACM on Web Conference 2025},
  pages={3183--3192},
  year={2025}
}

@inproceedings{park2024pacer,
  title={Pacer and runner: Cooperative learning framework between single- and cross-domain sequential recommendation},
  author={Park, Chung and Kim, Taesan and Yoon, Hyungjun and Hong, Junui and Yu, Yelim and Cho, Mincheol and Choi, Minsung and Choo, Jaegul},
  booktitle={Proceedings of the 47th International ACM SIGIR Conference on Research and Development in Information Retrieval},
  pages={2071--2080},
  year={2024}
}

@inproceedings{hou2022towards,
  title={Towards universal sequence representation learning for recommender systems},
  author={Hou, Yupeng and Mu, Shanlei and Zhao, Wayne Xin and Li, Yaliang and Ding, Bolin and Wen, Ji-Rong},
  booktitle={Proceedings of the 28th ACM SIGKDD conference on knowledge discovery and data mining},
  pages={585--593},
  year={2022}
}

@article{li2023diffurec,
  title={{DiffuRec}: A diffusion model for sequential recommendation},
  author={Li, Zihao and Sun, Aixin and Li, Chenliang},
  journal={ACM Transactions on Information Systems},
  volume={42},
  number={3},
  pages={1--28},
  year={2023},
  publisher={ACM New York, NY}
}

@article{xuan2024diffusion,
  title={Diffusion cross-domain recommendation},
  author={Xuan, Yuner},
  journal={arXiv preprint arXiv:2402.02182},
  year={2024}
}

@inproceedings{li2025cd,
  title={CD-CDR: Conditional diffusion-based item generation for cross-domain recommendation},
  author={Li, Hanyu and Li, Jiayu and Ma, Weizhi and Sun, Peijie and Wu, Haiyang and Wang, Jingwen and Yang, Yuekui and Zhang, Min and Ma, Shaoping},
  booktitle={Proceedings of the 48th International ACM SIGIR Conference on Research and Development in Information Retrieval},
  pages={1789--1798},
  year={2025}
}

@article{liu2024llm,
  title={{LLM-ESR}: Large language models enhancement for long-tailed sequential recommendation},
  author={Liu, Qidong and Wu, Xian and Wang, Yejing and Zhang, Zijian and Tian, Feng and Zheng, Yefeng and Zhao, Xiangyu},
  journal={Advances in Neural Information Processing Systems},
  volume={37},
  pages={26701--26727},
  year={2024}
}

@article{wang2025lecdsr,
  title={LeCDSR: Large language model enhanced cross-domain sequential recommendation},
  author={Wang, Shuliang and Zhu, Jiabao and Wang, Kaibo and Ruan, Sijie},
  journal={Information Fusion},
  pages={103762},
  year={2025},
  publisher={Elsevier}
}

@article{zhou2026d2tcdr,
  title={D2TCDR: Disentangled Diffusion-based Transfer for Cross-Domain Recommendation},
  author={Zhou, Hong and Lin, Xixun and Cao, Yanan and Zhu, Shichao and Jia, Renqi and Zhao, Xiangyu and Xu, Guandong and Guo, Li},
  journal={ACM Transactions on Information Systems},
  year={2026},
  publisher={ACM New York, NY}
}

@inproceedings{du2023frequency,
  title={Frequency enhanced hybrid attention network for sequential recommendation},
  author={Du, Xinyu and Yuan, Huanhuan and Zhao, Pengpeng and Qu, Jianfeng and Zhuang, Fuzhen and Liu, Guanfeng and Liu, Yanchi and Sheng, Victor S},
  booktitle={Proceedings of the 46th international ACM SIGIR conference on research and development in information retrieval},
  pages={78--88},
  year={2023}
}

@article{vaswani2017attention,
  title={Attention is all you need},
  author={Vaswani, Ashish and Shazeer, Noam and Parmar, Niki and Uszkoreit, Jakob and Jones, Llion and Gomez, Aidan N and Kaiser, {\L}ukasz and Polosukhin, Illia},
  journal={Advances in neural information processing systems},
  volume={30},
  year={2017}
}

@article{kingma2014adam,
  title={Adam: A method for stochastic optimization},
  author={Kingma, Diederik P and Ba, Jimmy},
  journal={arXiv preprint arXiv:1412.6980},
  year={2014}
}

@article{xin2025llmcdsr,
  title={{LLMCDSR}: Enhancing cross-domain sequential recommendation with large language models},
  author={Xin, Haoran and Sun, Ying and Wang, Chao and Xiong, Hui},
  journal={ACM Transactions on Information Systems},
  volume={43},
  number={5},
  pages={1--33},
  year={2025},
  publisher={ACM New York, NY}
}

@article{xu2025multi,
  title={A multi-view graph contrastive learning framework for cross-domain sequential recommendation},
  author={Xu, Zitao and Chen, Shu and Pan, Weike and Ming, Zhong},
  journal={ACM Transactions on Recommender Systems},
  volume={3},
  number={4},
  pages={1--28},
  year={2025},
  publisher={ACM New York, NY}
}

@article{boka2024survey,
  title={A survey of sequential recommendation systems: Techniques, evaluation, and future directions},
  author={Boka, Tesfaye Fenta and Niu, Zhendong and Neupane, Rama Bastola},
  journal={Information Systems},
  volume={125},
  pages={102427},
  year={2024},
  publisher={Elsevier}
}

@inproceedings{huang2024diit,
  title={DIIT: A Domain-Invariant Information Transfer Method for Industrial Cross-Domain Recommendation},
  author={Huang, Heyuan and Lou, Xingyu and Chen, Chaochao and Cheng, Pengxiang and Xin, Yue and He, Chengwei and Liu, Xiang and Wang, Jun},
  booktitle={Proceedings of the 33rd ACM International Conference on Information and Knowledge Management},
  pages={910--920},
  year={2024}
}

@inproceedings{ju2025revisiting,
  title={Revisiting self-attention for cross-domain sequential recommendation},
  author={Ju, Clark Mingxuan and Neves, Leonardo and Kumar, Bhuvesh and Collins, Liam and Zhao, Tong and Qiu, Yuwei and Dou, Qing and Nizam, Sohail and Yang, Sen and Shah, Neil},
  booktitle={Proceedings of the 31st ACM SIGKDD Conference on Knowledge Discovery and Data Mining V. 2},
  pages={1094--1105},
  year={2025}
}

@article{li2022adversarial,
  title={Adversarial learning for cross domain recommendations},
  author={Li, Pan and Brost, Brian and Tuzhilin, Alexander},
  journal={ACM Transactions on Intelligent Systems and Technology},
  volume={14},
  number={1},
  pages={1--25},
  year={2022},
  publisher={ACM New York, NY}
}

@inproceedings{lu2023contrastive,
  title={A contrastive learning framework for dual-target cross-domain recommendation},
  author={Lu, Jinhu and Sun, Guohao and Fang, Xiu and Yang, Jian and He, Wei},
  booktitle={Proceedings of the 31st ACM international conference on multimedia},
  pages={6332--6339},
  year={2023}
}

@inproceedings{liu2025flowcf,
  title={Flow Matching for Collaborative Filtering},
  author={Liu, Chengkai and Zhang, Yangtian and Wang, Jianling and Ying, Rex and Caverlee, James},
  booktitle={Proceedings of the 31st ACM SIGKDD Conference on Knowledge Discovery and Data Mining V. 2},
  pages={1765--1775},
  year={2025}
}

@inproceedings{liu2025fmrec,
  title={Flow matching based sequential recommender model},
  author={Liu, Feng and Zou, Lixin and Zhao, Xiangyu and Tang, Min and Dong, Liming and Luo, Dan and Luo, Xiangyang and Li, Chenliang},
  booktitle={Proceedings of the Thirty-Fourth International Joint Conference on Artificial Intelligence},
  pages={3108--3116},
  doi={10.24963/ijcai.2025/346},
  year={2025}
}

@article{li2025preference,
  title={Preference Trajectory Modeling via Flow Matching for Sequential Recommendation},
  author={Li, Li and Cheng, Mingyue and Ye, Yuyang and Liu, Zhiding and Chen, Enhong},
  journal={arXiv preprint arXiv:2508.17618},
  year={2025}
}

@inproceedings{li2025disco,
  title={DisCo: graph-based disentangled contrastive learning for cold-start cross-domain recommendation},
  author={Li, Hourun and Wang, Yifan and Xiao, Zhiping and Yang, Jia and Zhou, Changling and Zhang, Ming and Ju, Wei},
  booktitle={Proceedings of the AAAI Conference on Artificial Intelligence},
  volume={39},
  number={11},
  pages={12049--12057},
  year={2025}
}

@article{zhao2024domain,
  title={Domain-oriented knowledge transfer for cross-domain recommendation},
  author={Zhao, Guoshuai and Zhang, Xiaolong and Tang, Hao and Shen, Jialie and Qian, Xueming},
  journal={IEEE Transactions on Multimedia},
  volume={26},
  pages={9539--9550},
  year={2024},
  publisher={IEEE}
}

@inproceedings{zhang2025unveiling,
  title={Unveiling contrastive learning's capability of neighborhood aggregation for collaborative filtering},
  author={Zhang, Yu and Zhang, Yiwen and Zhang, Yi and Sang, Lei and Yang, Yun},
  booktitle={Proceedings of the 48th International ACM SIGIR Conference on Research and Development in Information Retrieval},
  pages={1985--1994},
  year={2025}
}

@inproceedings{xie2022contrastive,
  title={Contrastive learning for sequential recommendation},
  author={Xie, Xu and Sun, Fei and Liu, Zhaoyang and Wu, Shiwen and Gao, Jinyang and Zhang, Jiandong and Ding, Bolin and Cui, Bin},
  booktitle={2022 IEEE 38th international conference on data engineering (ICDE)},
  pages={1259--1273},
  year={2022},
  organization={IEEE}
}

@inproceedings{yu2022graph,
  title={Are graph augmentations necessary? simple graph contrastive learning for recommendation},
  author={Yu, Junliang and Yin, Hongzhi and Xia, Xin and Chen, Tong and Cui, Lizhen and Nguyen, Quoc Viet Hung},
  booktitle={Proceedings of the 45th international ACM SIGIR conference on research and development in information retrieval},
  pages={1294--1303},
  year={2022}
}

@article{zhao2023survey,
  title={A survey of large language models},
  author={Zhao, Wayne Xin and Zhou, Kun and Li, Junyi and Tang, Tianyi and Wang, Xiaolei and Hou, Yupeng and Min, Yingqian and Zhang, Beichen and Zhang, Junjie and Dong, Zican and others},
  journal={arXiv preprint arXiv:2303.18223},
  volume={1},
  number={2},
  pages={1--124},
  year={2023}
}

@article{wu2024survey,
  title={A survey on large language models for recommendation},
  author={Wu, Likang and Zheng, Zhi and Qiu, Zhaopeng and Wang, Hao and Gu, Hongchao and Shen, Tingjia and Qin, Chuan and Zhu, Chen and Zhu, Hengshu and Liu, Qi and others},
  journal={World Wide Web},
  volume={27},
  number={5},
  pages={60},
  year={2024},
  publisher={Springer}
}

@inproceedings{yuan2023go,
  title={Where to go next for recommender systems? id-vs. modality-based recommender models revisited},
  author={Yuan, Zheng and Yuan, Fajie and Song, Yu and Li, Youhua and Fu, Junchen and Yang, Fei and Pan, Yunzhu and Ni, Yongxin},
  booktitle={Proceedings of the 46th international ACM SIGIR conference on research and development in information retrieval},
  pages={2639--2649},
  year={2023}
}

@inproceedings{bao2023tallrec,
  title={TALLRec: An Effective and Efficient Tuning Framework to Align Large Language Model with Recommendation},
  author={Bao, Keqin and Zhang, Jizhi and Zhang, Yang and Wang, Wenjie and Feng, Fuli and He, Xiangnan},
  booktitle={Proceedings of the 17th ACM Conference on Recommender Systems},
  pages={1007--1014},
  year={2023},
  doi={10.1145/3604915.3608857}
}

@inproceedings{wei2024llmrec,
  title={LLMRec: Large Language Models with Graph Augmentation for Recommendation},
  author={Wei, Wei and Ren, Xubin and Tang, Jiabin and Wang, Qinyong and Su, Lixin and Cheng, Suqi and Wang, Junfeng and Yin, Dawei and Huang, Chao},
  booktitle={Proceedings of the 17th ACM International Conference on Web Search and Data Mining},
  pages={806--815},
  year={2024},
  doi={10.1145/3616855.3635853}
}

@inproceedings{hadad2025xcross,
  title={X-Cross: Dynamic Integration of Language Models for Cross-Domain Sequential Recommendation},
  author={Hadad, Guy and Roitman, Haggai and Eshel, Yotam and Shapira, Bracha and Rokach, Lior},
  booktitle={Proceedings of the 48th International ACM SIGIR Conference on Research and Development in Information Retrieval},
  pages={1497--1507},
  year={2025},
  doi={10.1145/3726302.3730117}
}

@inproceedings{yang2026pdflow,
  title={PDFlow: Popularity-Debiased Flow Matching for Sequential Recommendation},
  author={Yang, Zican and Li, Zeyu and He, Cong and Wu, Xiaotao and Liu, Guanfeng and Zhao, Pengpeng},
  booktitle={Proceedings of the Thirty-Fifth International Joint Conference on Artificial Intelligence},
  pages={3178--3186},
  year={2026},
  doi={10.24963/ijcai.2026/353}
}

@inproceedings{jin2026contrastive,
  title={Contrastive Flow Matching for Sequential Recommendation},
  author={Jin, Wangyu and Xue, Yang and Xia, Wenwen and He, Hongliang and Liu, Guanfeng and Zhao, Pengpeng},
  booktitle={Proceedings of the Thirty-Fifth International Joint Conference on Artificial Intelligence},
  pages={2744--2752},
  year={2026},
  doi={10.24963/ijcai.2026/305}
}

@inproceedings{liu2026bridging,
  title={Bridging Time and Domains: A Time-aware Framework for Cross-Domain Sequential Recommendation},
  author={Liu, Zemu and Qin, Zhida and Zhou, Pengzhan and Huang, Tianyu and Ding, Gangyi},
  booktitle={Proceedings of the ACM Web Conference 2026},
  pages={6091--6102},
  year={2026}
}

@inproceedings{qin2026bridging,
author = {Qin, Zhida and Liu, Zemu and Fu, Haoyan and Zhang, Chong and Huang, Tianyu and Li, Yidong and Ding, Gangyi},
title = {Bridging Behavior and Semantics for Time-aware Cross-Domain Sequential Recommendation},
booktitle = {Proceedings of the 49th International ACM SIGIR Conference on Research and Development in Information Retrieval},
pages = {1507–1518},
year={2026}
}

@article{liang2026semdiff,
  title={SemDiff: Semantic Guided Diffusion-Based Collaborative Filtering Framework},
  author={Liang, Xufeng and Qin, Zhida and Fu, Haoyan and Du, Enjun and He, Haotian and Zhang, Chong and Huang, Tianyu and Ding, Gangyi},
  journal={IEEE Transactions on Knowledge and Data Engineering},
  year={2026},
  publisher={IEEE}
}

@article{qin2024hyperbolic,
  title={Hyperbolic graph contrastive learning for collaborative filtering},
  author={Qin, Zhida and Cheng, Wentao and Ding, Wenxing and Ding, Gangyi},
  journal={IEEE Transactions on Knowledge and Data Engineering},
  volume={37},
  number={3},
  pages={1255--1267},
  year={2024},
  publisher={IEEE}
}


\end{document}